\documentclass[
10pt 
,leqno 
,english
]{article} 

\usepackage{my-preamble-article} 
\usepackage{my-macros} 

\usepackage{todonotes} 

\title{Ni-P coatings electroplating --- A review \\ Part I: Pure Ni-P alloy}

\author{Aleksandra \textsc{Lelević}}
\affil{\myAffiliation}

\begin{document}
\maketitle
\thispagestyle{empty}
\begin{abstract}
In the electroplating industry Ni-P coatings are extensively employed owing to their excellent properties which enable substrate protection against corrosion and wear. Depending on their composition and structure, as-plated deposits demonstrate good mechanical, tribological and electrochemical features, catalytic activity but also beneficial magnetic characteristics. With subsequent thermal treatment hardness of Ni-P metal-metalloid system can approach or be even higher than that of hard Cr coatings. The purpose of this paper is to provide a general survey of the research work dealing with the electrodeposition of Ni-P binary alloy coatings. Proposed phosphorus incorporation mechanisms, Ni-P alloy microstructure before and after thermal treatment, its mechanical, tribological, corrosion, catalytic and magnetic properties are considered, so are the key process variables influencing phosphorus content in the deposits and the roles of the main electrolytic bath constituents. Findings on the merits of employing pulse plating and fabrication of unconventional (layered and functionally graded) structures are succinctly explored. 
\end{abstract}
\begin{multicols}{2}
\section{Introduction}
In the modern world billions of dollars per year are lost due to machinery overhaul or complete breakdown related to material structural failures induced by progressive corrosion, wear, fatigue and rupture. Surface degradation impacts a large number of industrial sectors and can account for up to 3-4\% of the Gross Domestic Product in developed economies \cite{TottenLiang_surfaceModifMech_2004,Qin_micromanufactEnginTech_2015}. One way to address and diminish issues related to material deterioration is to impart on its surface a coating that will mitigate the influence of the surrounding environment and of the working conditions and prolong the service life of the equipment. 

For the deposition of coatings a variety of techniques exist that can be mechanical, physical, chemical and electrochemical in nature \cite{SidkyHocking_reviewInorgCoatandCoatProc1999}. 
Among them, simple electroplating offers many advantages \cite{LinLeeChen_ElectrodepNiPSulfamBath_2006,YuanSunYu_PrepOfAmorphNanocrNiP_2007,KnyazevFishgoitChernavskii_MagnPropOfEDAmorphNiP_2017}. 
It involves a process that takes place in the low cost electrolysis cell, in an aqueous solution, at normal pressure and at relatively low temperature, which makes it ideal for the industrial scaling up \cite{Paunovic_ModEP_2011,Kanani_EPPrincplProcPract_2005,Nasirpouri_EDNanostrMat_2017}. 
Particular appeal of electroplating is in the possibility of customizing appearance and properties of the coating by modifying the composition of the electroplating bath and/or the electroplating conditions and producing a wide range of metallic materials from metals and simple alloys to compositionally modulated systems and composites. Deposition rates in the order of several tens of micrometers per hour can be routinely achieved \cite{Bicelli_ReviewNanostrAspMetED_2008}. 

In terms of types of coatings fabricated through electroplating, in aerospace, automotive and general engineering industry functional Cr coatings are very widely employed owing to their remarkable combination of properties which include: high hardness, good corrosion, wear and heat resistance and low friction coefficient \cite{Paunovic_ModEP_2011}. However, the most common commercial Cr electrodeposition process employs an aqueous bath that contains hexavalent Cr. This ion is suspected to be a carcinogenic agent which in combination with its strong oxidizing nature imposes serious human health and environmental concerns \cite{LippmannMorton_EnvToxic_2008}. As a consequence, the use of hexavalent Cr has been limited by the European Union Restriction of Hazardous Substances Directive \cite{EC_theRoHSDirective}. In addition, the typical wetting agent-PFOS (perfluoro-octanesulfonic acid) employed in Cr electroplating to reduce the danger of the process through minimizing the misting generated during plating is being prohibited due to its exceptional stability which presents a problem for the environment \cite{Paunovic_ModEP_2011}. In the light of all mentioned, attention is focused on finding alternatives to Cr while maintaining the request for excellent functional properties of the fabricated deposits.

It has been reported by many authors that certain binary, ternary or quaternary alloy coatings may present a suitable replacement for hard Cr and may possess an interesting combination of features that can culminate even in a possible improvement in terms of performance characteristics \cite{MaWang_EDandCharactCoNiP_2013,Wang_NovEDGradNiPReplHardCr_2006a,Wang_CorrResLubrBehNovGradNiPReplHardCr_2006,Zhang_CoEDNiWDiamond_2016}. Amongst binary alloys Ni-P has attracted a lot of attention. Vast number of studies have been conducted which prove that coatings based on Ni-P with careful tailoring of composition and structure can offer smart and adaptive solutions to a wide range of environmental and working conditions \cite{Daly_EChemNiPAlloyForm_2003,BerkhZahavi_EDPropNiPandComp_1996}. Compared to Cr coatings obtained from hexavalent Cr baths, effluents resulting from Ni electroplating process have conventional, simpler treatment, such as for example processes called Clean Technologies (based on electrodialysis) \cite{DallaCosta_EvalElectrodialWastewat_2002}. Ni-P coatings exhibit many interesting features. They are, depending on the electroplating conditions and the applied post-treatment, characterized by: optimal mechanical properties, good wear and corrosion resistance, electro-catalytic activity and favourable tribological features. With proper customization of composition and structure and subsequent thermal treatment hardness of Ni-P electrodeposits can approach or surpass that of hard Cr coatings \cite{Nava_EffHeatTreatTribCorrNiP_2013}. Ni-P alloy electrodeposits find applications as protective, functional and decorative coatings primarily in automotive, aerospace and general engineering industries. Notable are the applications of Ni-P in the fabrication of decorative coatings in automotive industry, high precision parts, diffusion barriers, catalytic coatings for hydrogen evolution, thin film magnetic discs, microgalvanics, etc. \cite{Sadeghi_MicrStrEvolStrMechNiComp_2016}.

In order to attain the uttermost enhancement of the Ni-P deposit beneficial properties it is paramount to control the coating's composition and its microstructure. This can be achieved through altering electroplating bath composition, deposition conditions and by applying appropriate alloy post-treatment. To go beyond the limits of the properties that can be achieved for this binary alloy system, co-deposition of nano, sub-micron or micron size particles and fabrication of composite coatings is the centre of attention of the modern electroplating research and innovation efforts. While classical industrial demands on hard, homogeneous, wear resistant coatings and corrosion protection are still valid, added capacities in terms of: chemical stability, bio-compatibility, microstructured surfaces and functional coatings (lubricant, magnetic) are in high demand. This is where the production of composites but also compositionally modulated coatings, characterized by variable configuration across their thickness, can offer new possibilities and new combinations of properties that can be adapted to almost any need and satisfy the increasing demands on multi-functionality.
\section{Electroplating benefits and prospects}
Conventional electroplating possesses many advantages in terms of versatility, ease of use and cost effectiveness. The technique has many assets when compared to physical methods such as magnetron sputtering and chemical methods such as chemical vapour deposition. Physical vapour deposition is a high cost line-of-sight process with possibly poor throwing power, while in chemical vapour deposition high operating temperatures may cause substrate softening \cite{SidkyHocking_reviewInorgCoatandCoatProc1999,ScharfPrasad_SolidLubricantsReview_2013a}. 
Heat resistance of the substrate may very much limit the choice of the deposition technique, in particular for metals such as Al, Cu, Mg for which maximum deposition temperature is about \SI{100}{\degreeCelsius}.
Disadvantages of electroplating include mainly problems with current efficiency, throwing power, substrate adhesion and coating uniformity \cite{SidkyHocking_reviewInorgCoatandCoatProc1999}. 

In his work Dini \cite{Dini_PropCoatCompEpPvdCvdPs_1997} performed a comparison of coatings fabricated by various deposition techniques (PVD, CVD, PS and EP) on the bases of a wide range of properties, namely: structure, porosity, density, stress, corrosion, adhesion, tribology, fatigue, thermal conductivity, etc. His findings indicated that no coating technology provides superior results in terms of all considered parameters. The choice of coating technique must in fact be made by taking into consideration a number of factors, such as: substrate compatibility, adhesion, porosity, possibility of repair or re-coating, inter-diffusion, effect of thermal cycling, resistance to wear and corrosion, whilst simultaneously establishing a sensible balance between the obtained benefits and the incurred costs.

Electroplating possesses advantages over traditional electroless plating also, in terms of providing a higher deposition rate in a simpler electrolyte and the added benefit of the possibility to control deposit's composition and microstructure by changing the applied current waveform and electroplating conditions \cite{LinLeeChen_ElectrodepNiPSulfamBath_2006,YuanSunYu_PrepOfAmorphNanocrNiP_2007,Shi-EDNiCoMoS2_2008}. Thick deposits are easily achieved through electroplating. Amorphous Ni-P alloys can be electroplated up to mm thickness on widely differing substrates \cite{Daly_EChemNiPAlloyForm_2003}. However, it is still necessary to work on improving the electroplated deposits' uniformity and on maximizing the process efficiency when compared to electroless plating. By definition, electroless process must be meta-stable to ensure deposition. Plating in electroless mode is characterized by solution instability, slow deposition rate (generally around \SI{10}{\micro\metre\per\hour}, acidic electroless baths can plate at about \SI{25}{\micro\metre\per\hour}), high temperatures of operation (\SI{85}{\degreeCelsius} min for acceptable deposition rate), difficulties with barrel plating of small parts and high cost of the operation which is approximately \numrange{5}{10} times higher than for electrodeposition \cite{Luke_NiPElectroDep_1986}. Nevertheless, electroless plating has excellent throwing power and the advantage of producing deposits of uniform thickness particularly on components with complex shapes \cite{Kanani_EPPrincplProcPract_2004}. In order to achieve the latter in case of electroplating an intricate system of internal anodes and/or shielding is necessary due to the non-uniform current distribution characterizing this process. Ni-P deposits fabricated by electroless deposition are also reported to be harder and to possess better corrosion resistance compared to those obtained through electrodeposition \cite{Daly_EChemNiPAlloyForm_2003}.

Electroplating is quite a mature technique and although widely applied in industry since the end of the 19\textsuperscript{th} century, currently it is exhibiting stagnancy in terms of growth owing to the lack of technological innovations. However, according to several global market studies, electroplating is to demonstrate an expansion, predominantly in the Asia-Pacific excluding Japan region (APEJ). In the market report titled “Electroplating Market: Global Industry Analysis and Opportunity Assessment, 2016–2026” \cite{Electropl_MarRep}, Future Market Insights foresee that the global electroplating market is expected to expand at a compound annual growth rate of 3.7\% by the end of 2026 reaching worth of US\$ 21 Bn. APEJ will be according to them the fastest growing region and the Automotive and Electrical \& Electronics segments are estimated to collectively hold about 65\% of the total value share of the global electroplating market. On the basis of metal type, owing to the growth of the Electrical \& Electronics industry segments, copper and nickel are expected to gain significant basis points. As major growth drivers Future Market Insights identify applications across a diverse set of industries and expected eminent growth in the Asia Pacific region, while the biggest challenges according to them will be the rise of stern laws and stringent environmental regulations, decelerated economic growth in mature markets or markets in a state of equilibrium and the growing popularity of electroless nickel plating.
\section{Ni-P alloy electroplating}
\subsection{Phosphorus co-deposition mechanism}
\label{subsec:PIncorpMech}
Ni-P electroplating is predominantly performed in an aqueous electrolytic bath that contains Ni\textsuperscript{+2} ions as a source of nickel and a phosphorus oxyacid (or its salt) which acts as a source of phosphorus. Electrodeposition process is being driven by the electric current passing between the anode and the cathode (plating substrate) \cite{Kanani_EPPrincplProcPract_2004}. Standard reduction potentials for Ni (-0.25 V) and for P (-0.28 V) are near to each other facilitating their easy co-deposition. Earlier reviews that deal with the electrodeposition of Ni-P alloy and its composites were devised by Berkh and Zahavi \cite{BerkhZahavi_EDPropNiPandComp_1996} and by Daly and Barry \cite{Daly_EChemNiPAlloyForm_2003}. 

Two mechanisms are generally proposed in order to elucidate the process of phosphorus incorporation during Ni-P alloy electrodeposition, namely: direct and indirect mechanism.

In the direct mechanism it is proposed that Ni\textsuperscript{+2} ions and phosphorus oxyacid are directly reduced to Ni and P atoms which then form a Ni\textsubscript{x}P solid solution. According to Brenner \cite{Brenner_EDofQlloys_1963a} who first suggested this mechanism, phosphorus oxyacid is reduced directly to phosphorus and phosphorus is co-deposited with nickel owing to the polarization involved in the deposition of nickel assisting the deposition of phosphorus (induced co-deposition). The main argument against the direct deposition mechanism is the claimed impossibility of obtaining phosphorus in aqueous solutions through electrochemical methods in the absence of metal ions \cite{Crousier_EDniPAmorphMultilStr_1994}.

According to the indirect mechanism, first proposed by Fedotev and Vyacheslavov \cite{Fedotev_1959} and subsequently by Ratzker et al. \cite{Ratzker_EDndCorrPerNiP_1986}, phosphorus oxyacid is initially reduced to phosphine (PH\textsubscript{3}) in the presence of H\textsuperscript{+} ions, formed PH\textsubscript{3} is then oxidized to P in the presence of Ni\textsuperscript{+2} ions, while Ni\textsuperscript{+2} is simultaneously reduced to Ni. Zeller and Landau \cite{ZellerLandau_EDofNiP_1992} supported the hypothesis of PH\textsubscript{3} formation and suggested its subsequent reaction with Ni\textsuperscript{+2} producing Ni-P, H\textsuperscript{+} and a series of phosphorous oxyacids. 

Figure \ref{fig:phosph_incorp} gives an overview of the assumed half reactions describing phosphorus incorporation according to both proposed routes, direct and indirect. 
\begin{center}
	\captionsetup{type=figure}
	\includegraphics[width=\columnwidth]{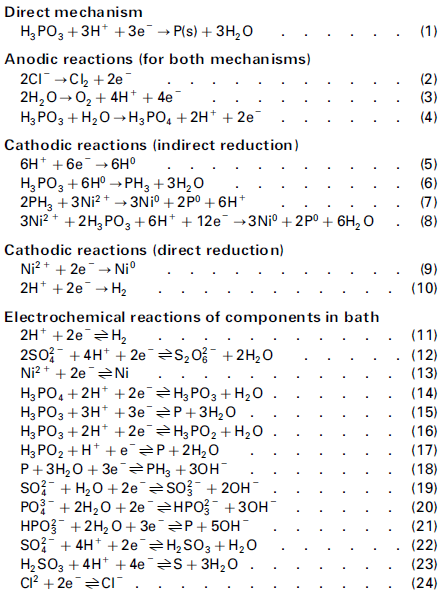}
	\captionof{figure}{Half reactions relevant to incorporation of phosphorus into Ni-P alloys. Reprinted with permission from \cite{Daly_EChemNiPAlloyForm_2003}. Copyright (2003) Taylor \& Francis. }
	\label{fig:phosph_incorp}
\end{center}

Occurrence of indirect mechanism has been corroborated by authors who have reported to have detected phosphine \cite{HarrisDang_MechOfPIncorp_1993,SaitouOkudikara_AmorphStrAndKinetOfPInc_2003,OrdineElectrochemStudNiPED_2006}. Harris and Dang \cite{HarrisDang_MechOfPIncorp_1993} quantified by chemical analysis PH\textsubscript{3} produced during Ni-P electrodeposition. Formation of this species was also identified by Crousier et al. \cite{Crousier_EDniPAmorphMultilStr_1994} who performed a cyclic voltammetry study. Zeng and Zhou \cite{ZengZhou_RamanStPinocrp_1999} obtained Raman spectra during Ni-P electrodeposition that indicated the formation of Ni(PH\textsubscript{3})\textsubscript{n} intermediate species. According to the authors, this intermediate was then oxidized by Ni\textsuperscript{+2} with the consequent formation of the alloy. 

Saitou et al. \cite{SaitouOkudikara_AmorphStrAndKinetOfPInc_2003} generated analytical solutions of kinetic equations for the reactions describing the phosphorus co-deposition according to both advocated mechanisms. The results predicted dependence of the phosphorous content in Ni-P electrodeposits on current density (Figure \ref{fig:Kin_P_Incorp}). Phosphorous content, as measured by the electron probe microanalyzer (EPMA), was found to decrease with the increase in current density which was in agreement with the solution of kinetic equations corresponding to the indirect mechanism of phosphorus co-deposition.

Morikawa and colleagues \cite{Morikawa_EDOfNiPFromNiCitr_1997} studied the electrodeposition of Ni-P from a citrate bath. Binding energy data obtained from X-ray photoelectron spectroscopy (XPS) indicated the direct reduction of H\textsubscript{2}PHO\textsubscript{3} to P, implying direct deposition of nickel-phosphide. Detection of phosphine by authors of previous studies Morikawa et al. explained with the possibility of partial reaction of phosphorus atoms on the electrode surface with hydrogen at low pH values, which in their case was negligible owing to high pH value of the employed citrate bath.
\begin{center}
	\captionsetup{type=figure}
	\includegraphics[width=0.8\columnwidth]{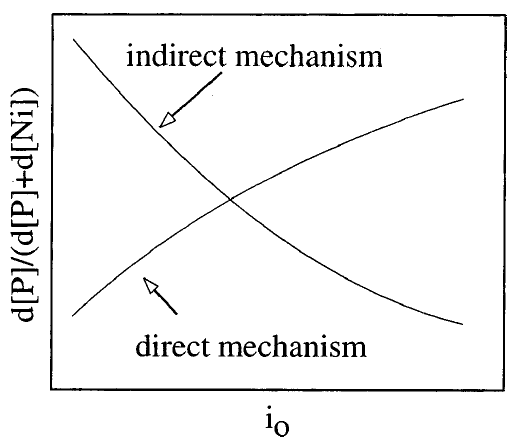}
	\captionof{figure}{Schematic diagram of phosphorous concentration vs. current density derived from rate equations for the indirect and direct mechanism. Reprinted with permission from \cite{SaitouOkudikara_AmorphStrAndKinetOfPInc_2003}. Copyright (2003) Electrochemical Society, Inc.}
	\label{fig:Kin_P_Incorp}
\end{center}

Ordine and coworkers \cite{OrdineElectrochemStudNiPED_2006} found kinetics of phosphorus incorporation to be strongly affected by the electrolytic solution pH value. They employed interfacial pH measurements, cathodic polarization curves and electrochemical impedance spectroscopy and observed different mechanisms of Ni and Ni-P electrodeposition at different degrees of solution acidity in the presence of NaH\textsubscript{2}PO\textsubscript{2}·H\textsubscript{2}O in the electroplating bath. 

Sotskaya and Dolgikh \cite{SotskayaDolgikh_KinetOfCathRedOfHypoph_2005} conducted electrochemical investigation of the kinetics regarding cathodic reduction of hypophosphite anion. Their results indicated that phosphorus formation proceeds via two parallel routes: electrochemical and chemical, whose realization depends on the nature of the employed metal catalyst. According to them atomic phosphorus is formed as a result of the cathodic reduction of hypophosphite ion. They observed also that the Ni-P alloy electro-synthesis takes place in the region of potentials, which does not correspond to the electro-reduction of hypophosphite anion, thus they concluded that a chemical disproportionation reaction of hypophosphite occurs in this case \cite{Savchuk_ExEffectofEsCondonNip_2017}.

Bredael et al. \cite{Bredael_JetCellRDElectrode_1993} proposed that two mechanisms are involved in the formation of Ni-P electrodeposits, one that is active at low current efficiency and leads to formation of amorphous deposits and the other dominating at high current efficiencies and leading to formation of crystalline ones. Such a claim was corroborated by polarization measurements in which a change of slope of the polarization curve was observed, this at the average current density corresponding to the transition from amorphous to crystalline structure.

One can infer that making an attempt to delineate the phosphorus co-deposition process with Ni constitutes a significant endeavour. Disagreements exists on whether direct or indirect mechanism can be assumed, with most of the authors supporting the latter. Additionally, indications exist that phosphorus incorporation is not completely straightforward and that several different mechanisms might govern its co-deposition depending on the process conditions. In the origin of the incurred difficulties lie the large number of factors directing and influencing the deposition process and many different approaches applied to define their nature and magnitude, all in various electroplating environments.
\subsection{Electrolytic bath composition}
The properties of the electrodeposited Ni-P coatings depend greatly on the composition of the employed electrolytic bath. The majority of nickel plating solutions are based on the ‘Watts’ formulation developed by Professor Oliver P. Watts in 1916 \cite{Watts_1916} owing to its simplicity and low cost. Modified Watts electrolyte for Ni-P plating combines the traditional nickel sulfate, nickel chloride and boric acid with a phosphorus oxyacid which is a source of phosphorus. Modified Watts bath containing sodium hypophosphite as a phosphorus source has also been reported in many works \cite{ZengZhou_RamanStPinocrp_1999,Hu_CompContrEPNiP_2001a,OrdineElectrochemStudNiPED_2006,SPYRELLIS_NiNi-PMatrixComp_2009,Dadvand_EDNiPfromCitrBath_2016}. Other than sulfate electrolyte, several other kinds of Ni-P aqueous electroplating baths exist, contingent on the nickel source nature mostly used are sulfamate and sulfonate ones.

\textit{Nickel source}--Most of the studies dedicated to nickel and nickel alloys electrodeposition have been restricted to simple sulfate containing or on Watts baths. A wide range of coatings can be deposited from these versatile and stable electrolytes, thus they still remain a basis for electroplating Ni-based coatings \cite{Walsh_SulfBathReview_2016}.

\begin{figure*}[t!]
	\centering
	\begin{subfigure}[t]{0.32\textwidth}
		\includegraphics[height=1.65in]{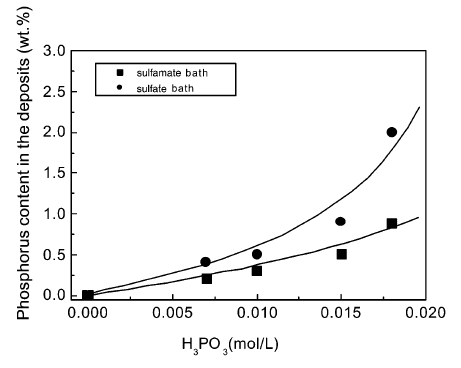}
		\caption{Phosphorus content vs C(H\textsubscript{3}PO\textsubscript{3})}
	\end{subfigure}
	~ 
	\begin{subfigure}[t]{0.32\textwidth}
		\includegraphics[height=1.5in]{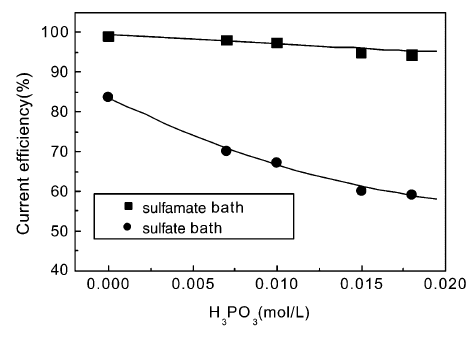}
		\caption{Current efficiency vs C(H\textsubscript{3}PO\textsubscript{3})}
	\end{subfigure}
	~ 
	\begin{subfigure}[t]{0.32\textwidth}
		\includegraphics[height=1.5in]{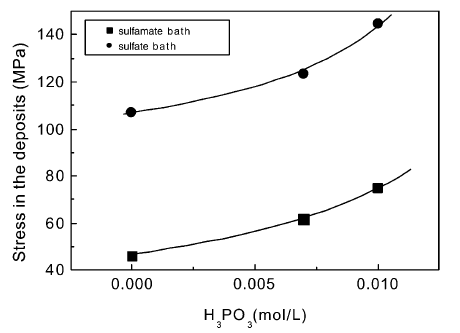}
		\caption{Stress in the deposit vs C(H\textsubscript{3}PO\textsubscript{3})}
	\end{subfigure}
	\caption{Comparison of stated parameters' values versus H\textsubscript{3}PO\textsubscript{3} concentration for the cases of Ni-P alloy electrodeposition from sulfamate and sulfate baths. Reprinted with permission from \cite{Seo_CharactNiPSulfamBath_2004}. Copyright (2003) Electrochemical Society, Inc.}
	\label{fig:Sulf_Sulfam_NiP}
\end{figure*}

Sulfamate baths are employed primarily for the purposes of high speed plating and electroforming. Nickel sulfamate possesses high solubility in aqueous solutions, thus higher nickel concentration can be achieved compared to other nickel electrolytes which facilitates higher plating rates \cite{DiBari_EDNi_2011,Baudrand_NiSulfamMystandPract_1996a}. Ni-P deposits obtained from sulfamate baths exhibit lower internal stress, good ductility and enable higher current efficiency \cite{Seo_CharactNiPSulfamBath_2004,Baudrand_NiSulfamMystandPract_1996a,LinLeeChen_ElectrodepNiPSulfamBath_2006}. Stress values of electrodeposits fabricated from sulfamate, sulfate and Watts baths are reported to be approximately 30, 180 and 250 MPa, respectively \cite{Seo_CharactNiPSulfamBath_2004}. Nevertheless, several issues need to be addressed when a nickel sulfamate bath is used. Sulfamate ion is stable in neutral or slightly alkaline solutions, however because of the nickel hydroxide precipitation these solutions are not used at pH values greater than 5. Sulfamate hydrolysis reaction has been found to proceed at an increased rate at lower pH values and at higher temperatures, thus sulfamate baths need to be operated at lower temperatures and higher pH values than Watts-type nickel plating solutions. pH values less than 3.0 and temperatures above \SI{70}{\degreeCelsius} ought to be avoided as nickel sulfamate can hydrolyze to its less soluble form-nickel ammonium sulfate \cite{DiBari_EDNi_2011,Baudrand_NiSulfamMystandPract_1996a}. Incorporation of ammonium and sulfate ions in the deposit can lead to the internal tensile stress augmentation. Sulfamate ion additionally tends to decompose at the anode; at insoluble anodes such as platinum and at passive nickel oxide electrodes, and its decomposition can result in intermediate species which may affect the quality of the electrodeposited coatings \cite{DiBari_EDNi_2011,Baudrand_NiSulfamMystandPract_1996a}. However, nickel sulfamate solutions are preferred in some cases over nickel sulfate solutions because of the superior mechanical properties of the fabricated coatings, high rates of deposition and lower influence of variations in pH and current density on the deposit's quality \cite{Martyak_EPLowStrNi_1998}. Figure \ref{fig:Sulf_Sulfam_NiP} shows the comparison of phosphorus content, current efficiency and stress in the obtained Ni-P deposits as a function of phosphorous acid concentration, for the sulfate and sulfamate electrolytes.

Another interesting electrolyte for Ni-P alloy electroplating is the one containing methanesulfonate or methanesulfonic acid. Sknar et al. \cite{Sknar_CharEdNiandNiPMethaneSulfon_2017} found the effect of reducing kinetic difficulties of the Ni-P alloy electrodeposition to be more pronounced when using methanesulfonate than when employing sulfate electrolyte. This is owing to weaker buffering properties and lower stability of nickel acido-complexes of methanesulfonate which contribute to the increase of concentration of the present electroactive species. Alloys obtained from sulfate baths possess higher phosphorus content which is explained by the elevated acidity in the near electrode layer due to stronger buffering properties of the sulfate electrolyte \cite{Savchuk_ExEffectofEsCondonNip_2017}. Methanesulfonic acid has good electrolytic conductivity and is capable of dissolving many metals as well as acting as a useful medium for dispersion of solids prior to electrophoretic coating. A diverse range of surface coatings and films are available from methanesulfonic acid electrolytes \cite{Walsh_VersEchCoatMethSulf_2014a}. Compared with known nickel plating baths, such as Watts bath, methanesulfonate bath can be considered to enable higher current density and result in higher throwing power. Maximum nickel deposition rates can be achieved with deposits having low porosity, low internal stress and high ductility \cite{Walsh_VersEchCoatMethSulf_2014a}. This organic acid may be considered as a ‘green’ electrolyte since it possesses few environmental, storage, transport or disposal problems being readily biodegradable. Additionally, methanesulfonic acid exhibits high solubility for metal salts such as those of Pb and Ag, which are soluble only in a limited number of acid electrolytes \cite{Gernon_EnvironmBenOfMethSulf_1999a}.

\textit{Phosphorus source}--Source of phosphorus in the Ni-P electroplating bath is typically a phosphorus oxyacid or its salt. There exist mainly two kinds of electrolyte systems for Ni-P electroplating: the one containing phosphorous acid and the other containing hypophosphite. Generally, the quality of Ni-P deposits electroplated from electrolyte containing hypophosphite is changeable because the deposits darken when electroplating time gets longer or current density becomes higher, while the Ni-P deposits electroplated from electrolyte containing phosphorous acid are smooth and bright \cite{YuanSunYu_PrepOfAmorphNanocrNiP_2007}. However, as reported, phosphine is more readily produced from hypophosphite than from phosphorous acid owing to the rate of reduction of H\textsubscript{3}PO\textsubscript{3} being the limiting factor in the phosphorous co-deposition in this case \cite{BerkhZahavi_EDPropNiPandComp_1996}. 

\textit{Anode dissolution}--If used as a contributing source of Ni in the bath, nickel chloride has two major functions: it appreciably increases solution conductivity thereby reducing voltage requirements and it is important in obtaining satisfactory dissolution of nickel anodes \cite{MacDougall_EffectofClion_1979,AbdElAal_AnodicDisofNi_2003}. Due to the increase in solution conductivity, plating thickness and cathodic current efficiency are reported to increase with the increase of chloride concentration \cite{Dadvand_EDNiPfromCitrBath_2016}. In the chloride electrolytes activity of Ni\textsuperscript{+2} ions is higher than in sulfate electrolyte and metal deposition potential is lower \cite{Mockute_EffOfClOnBehOfSacch_2001,Davalos_StofTheRoleofBorNiED_2013}. However, deposits obtained from chloride electrolytes with high chloride concentrations are reported to possess different texture and higher internal stress compared to those obtained from sulfate ones \cite{Davalos_StofTheRoleofBorNiED_2013}. It is possible to operate with zero chloride in the bath if sulfur activated nickel anodes are used under appropriate operating conditions, nonetheless a low concentration of nickel chloride (min \SI{5}{\gram\per\liter}) is generally recommended so as to provide reasonable insurance of satisfactory anode performance \cite{NiPlatingHandbook_2014}. Alternatively, nickel bromide that does not increase stress as much as chloride may be used.

\textit{Buffering}--As most authors claim, boric acid limits the effects on the solution pH value resulting from the discharge of hydrogen ions and simplifies pH control \cite{Ji_SurfpHMeasurNiED_1995}. However, many studies show that the beneficial influence of boric acid on Ni electrodeposition is somewhat complicated and that its effect stems further from simple buffering, extending to its impact on deposit's crystallographic structure, morphology, brightness, adhesion, etc. \cite{Davalos_StofTheRoleofBorNiED_2013}. Presence of boron in the bath waste is nonetheless reported to be harmful for the environment. Since the enforcement of Water Pollution Control Act in Japan \cite{Takuma_CompOfEnvImpNivsNewNi_2017} the imposed national minimum effluent standards have greatly influenced the electroplating industry. Thus, it became necessary to find more environmentally friendly solutions for the electrodeposition of nickel and its alloys. 

One of the possible alternatives is proposed in the terms of substituting boric acid with citric acid. This species can complex free nickel ions in the solution influencing the deposition rate but also acts as a buffering agent \cite{DoiMizumoto_NiPlCitrateBath_2001}. Doi et al. \cite{Doi_EffofpHNiCitrateEPBath_2004} found that the properties of Ni electrodeposits and resulting cathode current efficiency depend on the employed citrate bath pH. Electroplating resulted in high cathode current efficiency and hard deposits exhibiting crystal structure of nearly random orientation in case of the bath pH value being between 4 and 6, while when the pH value was 3.5 or less current efficiency was lower and hardness of the obtained deposits decreased. Dadvand and colleagues \cite{Dadvand_EDNiPfromCitrBath_2016} reported electrodeposition of Ni-P from a citrate bath containing sodium chloride, citric acid, nickel sulfate, ammonia and sodium hypophosphite. They found cathode current efficiency in low current density regions to be twice higher compared to the one achieved when employing modified Watts bath. Additionally, obtained deposits exhibited more uniform plating thickness. Even though complexes form between nickel and citrate ions and citrate ions as such can adsorb on the cathode surface, block active sights for Ni\textsuperscript{+2} discharge process and thus decrease the plating thickness, authors of the mentioned work estimated that this effect was small for citric acid concentrations below \SI{20}{\gram\per\liter}. Morikawa and colleagues \cite{Morikawa_EDOfNiPFromNiCitr_1997} employed Ni-citrate bath and obtained uniform and bright Ni-P electrodeposits with high throwing power. H\textsubscript{3}PO\textsubscript{3} was added in excess so as to generate a large amount of P atoms on the Ni electrode. Thus according to them, the bath composition was such that it satisfied the requirement necessary to form nickel phosphide with a stoichiometric composition equivalent to Ni\textsubscript{3}P in a wide range of current densities. Phosphorus content in the deposits was found to be constant around 25 at.\% in a broad range of plating conditions. 

When it comes to the environmental impact exerted by the electroplating process from either of the two baths: the one containing boric or the other containing citric acid, findings are quite contradictory. Takuma et al. \cite{Takuma_CompOfEnvImpNivsNewNi_2017} applied a life cycle assessment method to demarcate the extent of their influence in terms of human toxicity and ecotoxicity. Results indicated that the newly developed citrate plating bath exerts higher environmental impact compared to the traditional Watts electrolyte, this owing to the release of nickel chelated with citric acid whose harmful influence overshadows the benefit of reduced boron emissions achieved by substituting boric acid in the electroplating bath. 

In the Brenner type sulfate bath \cite{Brenner_EDofQlloys_1963a,Brenner_EDalloysPNiorCo_1950}, pH buffering is achieved generally by employing phosphoric acid (H\textsubscript{3}PO\textsubscript{4}) whose slight excess can moreover aid in the H\textsubscript{3}PO\textsubscript{3} oxidation prevention \cite{Pillai_EDNiPInDepthStPrepProp_2012a}. Pillai and colleagues \cite{Pillai_EDNiPInDepthStPrepProp_2012a} investigated electrodeposition of Ni-P alloy from a Brenner type bath containing both phosphorous and phosphoric acid. Their observation was that a 100\% nickel coating without any phosphorous incorporation was obtained when the plating was carried out in the absence of H\textsubscript{3}PO\textsubscript{3} indicating that phosphorous acid is the only electrochemically active phosphorous species which acts as a phosphorous source in the given solution.

Electroplating baths containing phosphorus and/or phosphoric acid can be very acidic, hence typically partial neutralization is necessary in order to increase the pH value of the solution up to an optimum value \cite{Brenner_EDalloysPNiorCo_1950}. For this purpose different alkaline compounds can be employed, i.e.: potassium or sodium carbonate, potassium or sodium hydroxide, ammonia. Dadvand et al. \cite{Dadvand_EDNiPfromCitrBath_2016} advocated the benefits of using ammonia for bath neutralization, as carbonate salts decrease the solution conductivity and alkali metal hydroxides promote the formation of nickel hydroxide complexes.

\textit{Additives}--Electrolyte additives are another crucial factor in the Ni-P electroplating process. They can be incorporated in the electrolytic solution in order to influence grain growth and crystallites orientation, to suppress undesirable secondary reactions, to promote the co-deposition of alloys, dissolution of metals, etc. Additives can be grouped according to their main purpose as: carriers, surface wetters, inhibitors or levellers, auxiliary brighteners, buffers, conductive electrolytes and chemical complexants \cite{Low_MultifNanostrCoatED_2015}. It is important to keep in mind that these species are also incorporated in the coating during its deposition, hence they may produce various adverse effects on the deposit's properties, for example decrease corrosion resistance and exacerbate mechanical properties. As such, their use should be carefully optimized. 

Grain refiners/carriers are usually aromatic organic compounds that contain sulfur whose co-deposition with nickel inhibits its grain growth. Most commonly employed grain refiner in Ni-P electrodeposition is saccharin. This species is reported to effectively reduce the internal stress of the fabricated deposits \cite{Chen_IntStressContrNiPPC_2010}, however its use comes with the concerns related to deterioration of the corrosion resistance due to incorporation of sulfur in the growing deposit \cite{Osaka_SulfurInflOnCorrProp_1999}. Saccharin is reported to prompt the decrease in the phosphorus content in the coating and to bring an improvement in current efficiency \cite{Hansal_PCEDniPSiC_2013,A.Zoikis-Karathanasis2015}. This is related to occupation of cathode surface active centres by saccharin molecules and its conversion which consumes available hydrogen, hence restricts the phosphorous co-deposition \cite{A.Zoikis-Karathanasis2015}.

It is known that properties of Ni electrodeposits depend on their crystallographic structure. Ni (100) soft-mode texture is associated with deposits that possess maximum ductility, minimum hardness and internal stress \cite{Pavlatou_HardEffIndSiCIncNiMatr_2006}. Textural modifications of this matrix can be induced by the introduction of certain additives \cite{A.Zoikis-Karathanasis2015}. Mixed crystal orientation can be achieved allowing to fabricate deposits with controlled properties.

Table \ref{table:NiP_bath_composition} gives an overview of the proposed compositions of the aqueous baths for Ni-P alloy electroplating. Applied deposition conditions and achieved phosphorus content where this information was available are stated along with some basic observations.

\textit{Non-aqueous baths}--All mentioned traditional Ni-P electroplating baths are corrosive and their use possesses a drawback in terms of generating toxic effluents. As an alternative to conventional electroplating solutions, You and colleagues \cite{You_NiPDES_2012} have proposed Ni-P electrodeposition from a bath containing choline chloride:ethylene glycol (1:2 molar ratio) deep eutectic solvent (DES) and NiCl\textsubscript{2}·6H\textsubscript{2}O and NaH\textsubscript{2}PO\textsubscript{2}·H\textsubscript{2}O as nickel and phosphorous sources, respectively. DESs are formed from mixtures of Lewis or Brønsted acids and bases and can contain a variety of anionic and/or cationic species. Employing ionic liquids in electroplating is beneficial owing to them being less corrosive, possessing a wider electrochemical window and exhibiting lower hydrogen evolution when compared to aqueous baths. As such they can present a suitable alternative for traditional plating solutions \cite{Neurohr_EDNiNonAqBaths_2015}. Disadvantages of DESs however include their lower electrical conductivity, metal salts solubility and higher viscosity in comparison with aqueous electrolytes.
{
\renewcommand{\arraystretch}{1.15}
\begin{table*}[htbp]
	\footnotesize
	\centering	
	\caption{Proposed bath compositions and deposition conditions for Ni-P alloy electroplating from sulfate, sulfamate and sulfonate baths according to various literature sources. Obtained phosphorus content is stated where applicable and some basic observations are noted. (jd-current density, t-temperature, $\varepsilon$-duty cycle, f-frequency, CE-cathode current efficiency)}
	\label{table:NiP_bath_composition}
	\begin{adjustbox}{height=0.92\textheight}
	\rotatebox{90}{%
		\begin{tabular}{cllcl}
			\toprule
		 \textit{Reference} &\textit{Bath composition} &\textit{Deposition conditions} &\textit{Phosphorus content} &\textit{Additional information} \\
			\hline\hline
		 &I: NiSO\textsubscript{4}·6H\textsubscript{2}O \SI{330}{\gram\per\liter} &pH 1.7-3.0 \\
		 &NiCl\textsubscript{2}·6H\textsubscript{2}O \SI{45}{\gram\per\liter} &\SI{63}{\degreeCelsius} \\
		 &H\textsubscript{3}BO\textsubscript{3} \SI{30}{\gram\per\liter} &2-\SI{5}{\ampere\per\square\deci\meter} &2-3 \%P \\
		 &H\textsubscript{3}PO\textsubscript{3} 0.225-\SI{2.25}{\gram\per\liter}\\
		 &Non-pitting agent \SI{0.15}{\gram\per\liter} \\
		 Durney, 1984 \cite{Durney_1984} \\
		 &II: NiSO\textsubscript{4}·6H\textsubscript{2}O \SI{150}{\gram\per\liter} &pH 0.5-1.0 \\
		 &NiCl\textsubscript{2}·6H\textsubscript{2}O \SI{45}{\gram\per\liter} &\SI{85}{\degreeCelsius} \\ 
		 &H\textsubscript{3}PO\textsubscript{4} \SI{50}{\gram\per\liter} &2-\SI{5}{\ampere\per\square\deci\meter} &12-15\%P \\
		 &H\textsubscript{3}PO\textsubscript{3} \SI{40}{\gram\per\liter} \\
		 \hline
		 &NiSO\textsubscript{4}·6H\textsubscript{2}O \SI{150}{\gram\per\liter} &&&Maximum P content from baths\\
		 &NiCl\textsubscript{2}·6H\textsubscript{2}O \SI{45}{\gram\per\liter} and:&&& I,II,III and IV is achieved for\\
		 &&&&following conditions: \\
		 &I: H\textsubscript{3}PO\textsubscript{4} \SI{50}{\gram\per\liter} &80 and \SI{90}{\degreeCelsius} & &I:35 wt.\%P for \SI{40}{\gram\per\liter} \\
		 &H\textsubscript{3}PO\textsubscript{3} 0-\SI{40}{\gram\per\liter} &$<$\SI{10}{\ampere\per\square\deci\meter} & & H\textsubscript{3}PO\textsubscript{3} and \SI{80}{\degreeCelsius} (CE$\sim$25\%) \\
		 & \\
		 Narayan and &II: H\textsubscript{3}PO\textsubscript{4} 0-\SI{200}{\gram\per\liter} &80 and \SI{90}{\degreeCelsius} &up to 35 wt.\%P &II:$\sim$20 wt.\%P for \SI{125}{\gram\per\liter} \\
		 Mungole, 1985 \cite{Narayan_EDNiPalloy_1985} &H\textsubscript{3}PO\textsubscript{3} \SI{20}{\gram\per\liter} &$<$\SI{10}{\ampere\per\square\deci\meter} & &H\textsubscript{3}PO\textsubscript{4} and \SI{80}{\degreeCelsius} (CE$\sim$50\%) \\
		 & \\
		 &III: H\textsubscript{3}PO\textsubscript{4} \SI{50}{\gram\per\liter} &\SI{90}{\degreeCelsius} & &III:$\sim$30 wt.\%P for \SI{12}{\gram\per\liter} \\
		 &H\textsubscript{3}PO\textsubscript{3} \SI{20}{\gram\per\liter}&$<$\SI{10}{\ampere\per\square\deci\meter} & & carbonate (CE$\sim$70\%) \\
		 &NiCO\textsubscript{3}·NiOH\textsubscript{2}·4H\textsubscript{2}O 0-\SI{15}{\gram\per\liter} & \\
		 & \\
		 &IV: H\textsubscript{3}PO\textsubscript{4} \SI{50}{\gram\per\liter} &70-\SI{90}{\degreeCelsius} & &IV:30 wt.\%P for \SI{5}{\ampere\per\square\deci\meter} \\
		 &H\textsubscript{3}PO\textsubscript{3} \SI{20}{\gram\per\liter} & 5-\SI{40}{\ampere\per\square\deci\meter}& & and \SI{70}{\degreeCelsius} (CE$\sim$25\%) \\
		 \hline
		 &NiSO\textsubscript{4}·6H\textsubscript{2}O \SI{150}{\gram\per\liter} &pH 0.43-1.0 & & Transition to amorphous \\
		 Bredael et al., 1993 \cite{Bredael_JetCellRDElectrode_1993} &NiCl\textsubscript{2}·6H\textsubscript{2}O \SI{50}{\gram\per\liter} &\SI{60}{\degreeCelsius} &up to 20 wt.\%P & structure at $\geq$12 wt.\% P\\
		 &H\textsubscript{3}PO\textsubscript{4} \SI{42.5}{\gram\per\liter} &2-\SI{150}{\ampere\per\square\deci\meter} & & \\
		 &H\textsubscript{3}PO\textsubscript{3} 3-\SI{70}{\gram\per\liter} &RDE 750 rev/min\\
		 \hline
		 &NiSO\textsubscript{4} 0.38M &pH 1.5-4.5 & &Maximum P content 25 at.\%\\
		 &NiCl\textsubscript{2} 0.13M &30-\SI{90}{\degreeCelsius} &0-25 at.\%P &for $\sim$ 0.5M H\textsubscript{3}PO\textsubscript{3}, at \SI{60}{\degreeCelsius}, \\
		 Morikawa et al., 1996 \cite{Morikawa_EDOfNiPFromNiCitr_1997} &H\textsubscript{3}BO\textsubscript{3} 0.49M &1-\SI{30}{\ampere\per\square\deci\meter} & &\SI{3}{\ampere\per\square\deci\meter} and pH 3.5 \\
		 &Citric acid 0.5M & & &(CE$\sim$25\%);\\
		 &H\textsubscript{3}PO\textsubscript{3} 0-2M & & &For H\textsubscript{3}PO\textsubscript{3} $>$0.5M P levels off \\
		 \bottomrule
		\end{tabular}%
	}
\end{adjustbox}
\end{table*}
}
{
	\renewcommand{\arraystretch}{1.15}
\begin{table*}[htbp]
	\footnotesize
	\centering
		\begin{adjustbox}{height=0.92\textheight}
	\rotatebox{90}{%
		\begin{tabular}{cllcl}
			\toprule
			\textit{Reference} &\textit{Bath composition} &\textit{Deposition conditions} &\textit{Phosphorus content} &\textit{Additional information} \\
			\hline\hline
			&NiSO\textsubscript{4}·6H\textsubscript{2}O \SI{330}{\gram\per\liter} &pH 1.0 and 4.0 &13-28 at.\%P (pH 1.0 &Composition control by \\
			Hu and Bai, 2001 \cite{Hu_CompContrEPNiP_2001a} &NiCl\textsubscript{2} \SI{45}{\gram\per\liter} & 20 and \SI{50}{\degreeCelsius} &, NaH\textsubscript{2}PO\textsubscript{2}·H\textsubscript{2}O 1M) &simultaneous change\\
			Hu and Bai, 2003 \cite{Hu_InfPContPhysChemPropNiP_2003}&H\textsubscript{3}BO\textsubscript{3} \SI{37}{\gram\per\liter} &5 and \SI{25}{\ampere\per\square\deci\meter} &4-12 at.\%P (pH 4.0 &of t and jd; Amorphous\\
			&NaH\textsubscript{2}PO\textsubscript{2}·H\textsubscript{2}O 0.5 and 1M &200 and 400 rev/min &, NaH\textsubscript{2}PO\textsubscript{2}·H\textsubscript{2}O 0.5M) & structure for $\geq$12 at.\%P\\
			\hline
			&NiSO\textsubscript{4}·6H\textsubscript{2}O \SI{150}{\gram\per\liter} &pH 1.8 &&Primary nucleation of Ni \\
			&NiCl\textsubscript{2}·6H\textsubscript{2}O \SI{45}{\gram\per\liter} &\SI{55}{\degreeCelsius} &27 at.\%P &followed by Ni-P formation; \\
			Kurowski et al., 2002 \cite{Kurowski_InStagNiPED_2002} &H\textsubscript{3}PO\textsubscript{4} \SI{50}{\gram\per\liter} &-420 mV (SHE) &&Growth driven by\\
			&NiCO\textsubscript{3} \SI{40}{\gram\per\liter} &&&surface diffusion of \\
			&H\textsubscript{3}PO\textsubscript{3} \SI{40}{\gram\per\liter} &&&Ni and P species\\
			\hline
			&NiCl\textsubscript{2}·6H\textsubscript{2}O 0.2M &pH 3.3 & &10 at.\%P for 4M NH\textsubscript{4}Cl\\
			Li et al., 2003 \cite{Li_EffAmmonLowTempEDNiP_2003} &NaH\textsubscript{2}PO\textsubscript{2}·H\textsubscript{2} 0.1M &\SI{25}{\degreeCelsius} &2-10 at.\%P &(CE$\sim$60\%) \\
			&NH\textsubscript{4}Cl 0.5-4M & 5-\SI{50}{\ampere\per\square\deci\meter}\\
			\hline
			&NiSO\textsubscript{4}·6H\textsubscript{2}O \SI{137}{\gram\per\liter} &pH 1.5 & &Nanocrystalline deposit; \\
			&NiCO\textsubscript{3} \SI{36.5}{\gram\per\liter} &\SI{70}{\degreeCelsius} &0.5-2.5 wt.\%P &Maximum hardness (990HV) \\
			Jeong et al., 2003 \cite{Jeong_RelHardAbrWearEDNiP_2003} &H\textsubscript{3}PO\textsubscript{3} 2-\SI{3}{\gram\per\liter} & 1-\SI{3}{\ampere\per\square\deci\meter} & &and wear resistance after\\
			& Saccharin \SI{5}{\gram\per\liter} &Magnetic stirring & &heat treatment at \SI{370}{\degreeCelsius}\\
			& SLS \SI{0.1}{\gram\per\liter} \\
			\hline
			&Ni(SO\textsubscript{3}NH\textsubscript{2})\textsubscript{2}·4H\textsubscript{2}O &pH 1.0-2.5 & &Maximum P content 14 wt.\%\\
			&(Ni\textsuperscript{+2} \SI{90}{\gram\per\liter}) &\SI{50}{\degreeCelsius} &2-12 wt.\%P & for $\varepsilon$=0.1 and f$>$100Hz\\
			Lin et al., 2006 \cite{LinLeeChen_ElectrodepNiPSulfamBath_2006} &NiCl\textsubscript{2}·6H\textsubscript{2}O \SI{3}{\gram\per\liter} & jp \SI{8}{\ampere\per\square\deci\meter} & &(CE$\sim$ 80\%);\\
			Chen et al., 2010 \cite{Chen_IntStressContrNiPPC_2010} &H\textsubscript{3}BO\textsubscript{3} \SI{40}{\gram\per\liter} &10-500 Hz & &Deposit still crystalline;\\
			&H\textsubscript{3}PO\textsubscript{3} \SI{10}{\gram\per\liter} &$\varepsilon$ 0.1-0.5 & &Compressive internal stress\\
			&SDS \SI{0.4}{\gram\per\liter} &Air bubbling & &(-40 MPa)\\
			\hline
			&NiSO\textsubscript{4}·6H\textsubscript{2}O \SI{240}{\gram\per\liter} &pH 0.8-1.8 & &Maximum P content \\
			&NiCl\textsubscript{2}·6H\textsubscript{2}O \SI{28}{\gram\per\liter} &45 and \SI{75}{\degreeCelsius} &up to 10 wt.\%P &for higher H\textsubscript{3}PO\textsubscript{3}  \\
			Yuan et al., 2007 \cite{YuanSunYu_PrepOfAmorphNanocrNiP_2007} &H\textsubscript{3}BO\textsubscript{3} \SI{30}{\gram\per\liter} &4 and \SI{10}{\ampere\per\square\deci\meter} & &concentration and\\
			&H\textsubscript{3}PO\textsubscript{3} 2 and \SI{8}{\gram\per\liter} &Magnetic stirring & &lower pH values\\
			& &150 and 550 rev/min \\
			\hline	
			&Ni(SO\textsubscript{3}NH\textsubscript{2})\textsubscript{2}·4H\textsubscript{2}O &pH 1.5-3.5 & &Maximum P content \\
			&(Ni\textsuperscript{+2} \SI{96}{\gram\per\liter}) &\SI{45}{\degreeCelsius} & &for higher H\textsubscript{3}PO\textsubscript{3}  \\
			Chang et al., 2008 \cite{Chang_EDNiPSulfamRelpHStr_2008} &NiCl\textsubscript{2}·6H\textsubscript{2}O \SI{4}{\gram\per\liter} &\SI{2}{\ampere\per\square\deci\meter} &1-4.5 wt.\%P &concentration and \\
			&H\textsubscript{3}BO\textsubscript{3} \SI{32}{\gram\per\liter} &Air bubbling & &lower pH values \\
			&H\textsubscript{3}PO\textsubscript{3} 0-\SI{4}{\gram\per\liter} & & &(CE$\sim$ 70\%) \\
			&Wetting agent \SI{0.3}{\milli\liter\per\liter} \\
			\bottomrule
		\end{tabular}%
	}
\end{adjustbox}
\end{table*}
}
{
	\renewcommand{\arraystretch}{1.15}
\begin{table*}[htbp]
	\footnotesize
	\centering
	\begin{adjustbox}{height=0.92\textheight}
	\rotatebox{90}{%
		\begin{tabular}{cllcl}
			\toprule
			\textit{Reference} &\textit{Bath composition} &\textit{Deposition conditions} &\textit{Phosphorus content} &\textit{Additional information} \\
			\hline\hline
			&NiSO\textsubscript{4}·6H\textsubscript{2}O 170 or \SI{330}{\gram\per\liter} &pH 0.5-3.0 \\
			Schlesinger and &NiCl\textsubscript{2}·6H\textsubscript{2}O 35-\SI{55}{\gram\per\liter} &60-\SI{95}{\degreeCelsius}\\
			Paunovic, 2010 \cite{Paunovic_ModEP_2011}&H\textsubscript{3}BO\textsubscript{3} 0 or \SI{40}{\gram\per\liter} &2-\SI{5}{\ampere\per\square\deci\meter}\\
			&H\textsubscript{3}PO\textsubscript{4} 50 or \SI{0}{\gram\per\liter} \\
			&H\textsubscript{3}PO\textsubscript{3} 2-\SI{40}{\gram\per\liter} \\
			\hline
			&NiSO\textsubscript{4}·7H\textsubscript{2}O \SI{150}{\gram\per\liter} &pH$\sim$1.5 & &P content decreases with \\
			&NiCl\textsubscript{2}·6H\textsubscript{2}O \SI{45}{\gram\per\liter} &50-\SI{80}{\degreeCelsius} &0-20 wt.\%P &H\textsubscript{3}PO\textsubscript{3} concentration decrease\\
			&H\textsubscript{3}PO\textsubscript{4} 0-\SI{40}{\gram\per\liter} &5-\SI{30}{\ampere\per\square\deci\meter} & &and increase of jd and t\\
			Pillai et al.,2012 \cite{Pillai_EDNiPInDepthStPrepProp_2012a} &H\textsubscript{3}PO\textsubscript{3} 0-\SI{20}{\gram\per\liter} & & &(up to \SI{15}{\gram\per\liter}H\textsubscript{3}PO\textsubscript{3});\\
			&SLS \SI{0.25}{\gram\per\liter}  & & &Deposit amorphous for $\geq$9 wt.\%P;\\
			& & & &Hardness 8.57 GPa (4-7 wt.\%P),\\
			& & & &after annealing at \SI{400}{\degreeCelsius}\\
			& & & &maximal hardness (12 GPa)\\
			\hline
			&NiSO\textsubscript{4}·6H\textsubscript{2}O 0.65M &pH 1.5 & &Transformation to crystalline\\
			&NiCl\textsubscript{2}·6H\textsubscript{2}O 0.75M & &10.6 at.\%P &at 500-\SI{600}{\degreeCelsius};\\
			Nava et al., 2013 \cite{Nava_EffHeatTreatTribCorrNiP_2013} &H\textsubscript{3}BO\textsubscript{3} 0.15M &\SI{3}{\ampere\per\square\deci\meter} & &Annealed deposit hardness\\
			&NaCl 2M & & &990 HV\\
			&H\textsubscript{3}PO\textsubscript{3} 0.1M \\
			\hline
			&NiSO\textsubscript{4}·6H\textsubscript{2}O 0.2M &pH 3.0-4.0 &at surface: &Soft magnetic character, \\
			&H\textsubscript{3}BO\textsubscript{3} 0.005M &\SI{70}{\degreeCelsius} &$\sim$8-12 at.\%P &plating conditions dependant;  \\
			Alleg et al., 2016 \cite{Alleg_MicrostrAndMagnPropNiP_2016} &NaH\textsubscript{2}PO\textsubscript{2}·H\textsubscript{2}O 0.1M &-1.3 to -1 V &across depth: &Deposits mixtures of  \\
			&Saccharin 0.005M & &$\sim$4-6 at.\%P &Ni(P) solid solutions and  \\ 
			&NaCl 0.7M  &&&amorphous or Ni\textsubscript{2}P phase\\
			\hline
			&I: Ni(CH\textsubscript{3}SO\textsubscript{2})\textsubscript{2} 1M &pH 3.0 & &Max P content for 0.12M \\
			&NaCl 0.3M &\SI{60}{\degreeCelsius} & &NaH\textsubscript{2}PO\textsubscript{2} and \SI{2}{\ampere\per\square\deci\meter}: \\
			&NaH\textsubscript{2}PO\textsubscript{2} 0.03-0.12M &2-\SI{7}{\ampere\per\square\deci\meter} & &$\sim$8 (I) and $\sim$13 wt.\%P (II); \\
			Sknar et al., 2017 \cite{Sknar_CharEdNiandNiPMethaneSulfon_2017} & & &up to 13 wt.\%P \\
			&II: NiSO\textsubscript{4} 1M & & &P content higher from \\
			&NaCl 0.3M & & &sulfate bath;\\
			&H\textsubscript{3}BO\textsubscript{3} 0.7M & & &Internal stress and hardness \\
			&NaH\textsubscript{2}PO\textsubscript{2} 0.03-0.12M & & &better from sulfamate \\
			\bottomrule
		\end{tabular}%
	}
\end{adjustbox}
\end{table*}
}

It has been proposed that environmental issues related to the use of aqueous baths in Nickel electroplating can be mitigated also by performing electrodeposition in a supercritical CO\textsubscript{2} fluid \cite{Yoshida_EPNissCO2_2002,Yoshida_CO2EPNi2003}. Chuang and colleagues \cite{Chuang_EffSurfEDNiPssCO2_2013} studied the properties and the electrodeposition behaviour of Ni-P coatings in emulsified supercritical CO\textsubscript{2} in the presence of suitable surfactants. They observed improved hardness, wear resistance and surface quality of the obtained coatings compared to deposits fabricated from conventional aqueous electroplating solutions. However, current efficiency from supercritical CO\textsubscript{2} bath was lower.
\subsection{Current efficiency}
In accordance with both proposed phosphorus incorporation mechanisms, direct or indirect, concurrent reduction of protons at the surface of the growing deposit layer is necessary for the co-deposition of phosphorus with nickel (Figure \ref{fig:phosph_incorp}). However, subsequent recombination of formed H\textsubscript{ads} limits the amount of phosphorus that can be deposited. Thus, surface H\textsuperscript{+} concentration affects both the phosphorus content and the cathode current efficiency whose magnitude can never reach values close to 100\% as long as the phosphorus is co-deposited with nickel.

Cathode current efficiency of Ni-P electroplating is primarily affected by the electrolytic bath composition, plating regime, applied current density, temperature, bath pH value and agitation rate. Ross et al. \cite{ROSS_EDtechniqMiltLay_1993} found that the current efficiency decreases as the deposit's phosphorus content increases and that it depends on the rotation speed of the cathode. In the case of rotating-disc electrode (RDE), maximum efficiency was attained for low phosphorus plating solutions and low rotation speeds. In general, electrodeposition of high phosphorus Ni-P coatings proceeds at a lower current efficiency than the electrodeposition of low phosphorus ones \cite{Bredael_JetCellRDElectrode_1993,LinLeeChen_ElectrodepNiPSulfamBath_2006}. Naryan and Mungole \cite{Narayan_EDNiPalloy_1985} found that the current efficiency decreases with increasing H\textsubscript{3}PO\textsubscript{3} concentration and decreasing temperature of the sulfate electroplating bath. Similarly, Seo and colleagues \cite{Seo_CharactNiPSulfamBath_2004} observed that with increasing H\textsubscript{3}PO\textsubscript{3} concentration in the sulfamate bath, phosphorus content in the deposit increases, current efficiency decreases and stress in the obtained deposits augments. Morikawa \cite{Morikawa_EDOfNiPFromNiCitr_1997} observed that in citrate bath current efficiency exhibits a maximum at intermediate H\textsubscript{3}PO\textsubscript{3} concentrations.

Cathode current efficiency in Ni-P electroplating is generally found to increase with the increase of current density \cite{Low_EDCompCoatNPDep_2006a,Toth-Kadar_PrepCharNiPAmorph_1987,Luke_NiPElectroDep_1986}, bath temperature \cite{BerkhZahavi_EDPropNiPandComp_1996} and its pH value \cite{Goldman_ShortvwCompMod_1986,BerkhZahavi_EDPropNiPandComp_1996}. Luke \cite{Luke_NiPElectroDep_1986} found that the cathode current efficiency does not vary appreciably with current density in the higher range (above $\sim$\SI{20}{\ampere\per\square\deci\metre}) but increases with the increase of current density at its lower values. Similar trend was observed by Toth-Kadar et al. \cite{Toth-Kadar_PrepCharNiPAmorph_1987}. Li et al. \cite{Li_EffAmmonLowTempEDNiP_2003} found cathode current efficiency to be lower for a larger total current density. This is owing to the discharge reaction of H\textsuperscript{+} ions which results in the evolution of hydrogen gas bubbles. This reaction shares a larger portion of the electrodeposition current at higher total current densities. 

In general, reactions involved in the conversion of phosphorus in the solution into phosphorus in the deposit can be interpreted differently and thus the discrepancies in elaborating cathode current efficiency behaviour may arise. 

Cathode current efficiency of Ni-P alloy electrodeposition can be as reported significantly improved by employing pulse instead of direct current plating \cite{LinLeeChen_ElectrodepNiPSulfamBath_2006}. Benefits of using pulse plating for Ni-P deposits fabrication are explored in Section \ref{Pulse plating}.

As mentioned in the previous section, certain additives such as saccharin are found to improve the current efficiency of Ni-P electrodeposition \cite{A.Zoikis-Karathanasis2015}.

In Ni-P electroplating, anodes are usually made of nickel and current efficiency of nickel dissolution in additive free solution is approximately 100\% (small percentage of the current is consumed on the discharge of hydrogen ions from water). Hence there exists a difference in cathode and anode efficiency which leads to nickel ions build up in the solution and its pH value increase. This problem can be solved by solution drag-out \cite{DiBari_NiPlating_2001} and by the use of soluble and inert anodes connected to different power supplies, while the current to each is controlled in order to compensate for the low cathode efficiency and incurred drag-out losses \cite{Luke_NiPElectroDep_1986,Paunovic_ModEP_2011}.
\subsection{Crystallographic structure}
\begin{figure*}[htb!]
	\centering    
	\includegraphics[height=120mm]{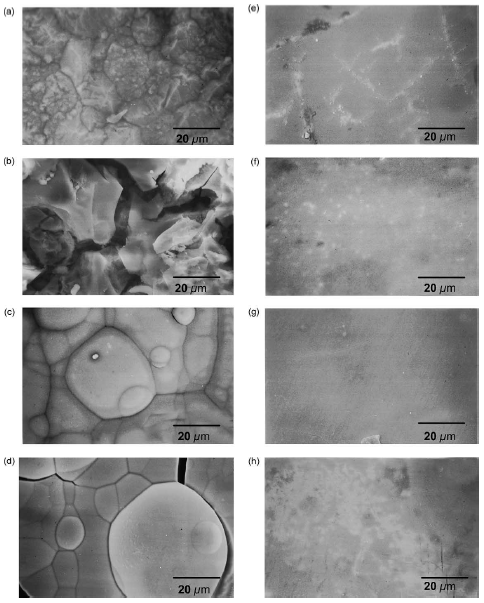}
	\caption{SEM photographs of Ni-P deposits with P contents of (a) 0, (b) 4, (c) 8, (d) 12, (e) 17, (f) 20, (g) 24 and (h) 28 at.\%. Reprinted with permission from \cite{Hu_InfPContPhysChemPropNiP_2003}. Copyright (2003) Elsevier.}
	\label{fig:NiP_SEM}
\end{figure*}

Ni-P electrodeposits' crystallographic structure is strongly influenced by their phosphorus content and the employed deposition conditions. Incorporation of even small amounts of phosphorous in the nickel lattice substantially refines the nickel grains. In the case of low-phosphorus Ni-P electrodeposits, obtained XRD patterns reveal a set of diffraction peaks corresponding to face-centred cubic (f.c.c.) nickel, i.e.: (111), (200), (220), (311), (222) \cite{Apachitei_AutocatNionAl_2000}. Presence of these textures is an indication of crystallinity of the low-phosphorus coatings in which case a material is a supersaturated solution of phosphorus in f.c.c. nickel \cite{ROSS_EDtechniqMiltLay_1993}. As the phosphorus content in Ni-P deposits increases, (111) reflection becomes broader while others disappear, as it can be seen on Figure \ref{fig:XRD_NiP}. This is indicative of nickel crystalline structure losing its long-range order due to the increased difficulty of more phosphorus atoms accommodating into the nickel lattice. Co-deposition of phosphorus in octahedral interstial sites of f.c.c. nickel inhibits the surface diffusion of Ni atoms and the subsequent crystal growth \cite{Daly_EChemNiPAlloyForm_2003}. As the phosphorus content in the deposit increases the nucleation becomes dominant over nucleus growth and when the critical phosphorus content is exceeded amorphous structure characterized by a short range order of only few atomic distances is obtained. More colony-like morphology is achieved where each colony consists of numerous grains with smaller grain size thereby making the coating brighter and smoother in appearance \cite{Pillai_EDNiPInDepthStPrepProp_2012a}. Figure \ref{fig:NiP_SEM} shows SEM photographs of Ni-P deposits with varying phosphorus contents.
\begin{center}
	\captionsetup{type=figure}
	\includegraphics[width=0.9\columnwidth]{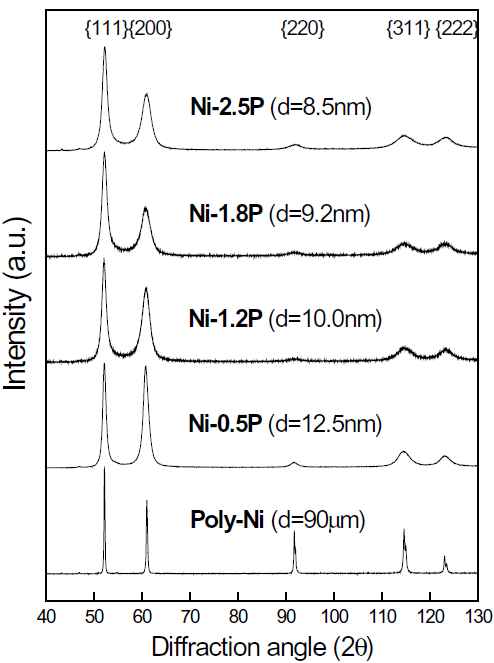}
	\captionof{figure}{XRD patterns of polycrystalline Ni and nanocrystalline Ni-P coatings. Reprinted with permission from \cite{Jeong_RelHardAbrWearEDNiP_2003}. Copyright (2003) Elsevier.}
	\label{fig:XRD_NiP}
\end{center}

At high-phosphorus content XRD pattern of a Ni-P electrodeposit exhibits only a diffuse broad peak. Hence, an increase in phosphorus content converts the microstructure of Ni-P alloys from crystalline to amorphous, which results in the decreased ductility and the increased corrosion resistance. It has been hypothesized that adsorbed hydrogen blocks the surface and prevents regular crystal growth, hence playing a crucial factor in obtaining N-P deposits with amorphous structure, however amorphous deposits can be fabricated with high current efficiencies and low hydrogen evolution thus indicating that the adsorbed hydrogen does not play a major role \cite{Daly_EChemNiPAlloyForm_2003}. 

Bredael et al. \cite{Bredael_AmorpCrystEDNiP_1994} observed significant broadening of the (111) XRD peak for Ni-P deposits containing more than 9 wt.\% of phosphorus. However they stated that the observed effect, other than being a sign of the structure becoming amorphous, can also originate from a crystalline material with very fine grains (1nm) and may be due to non-uniform internal stresses and stacking faults. Hence, as they asserted, it is not possible to exactly distinguish between amorphous and microcrystalline structure based only on XRD patterns. Nonetheless, conventionally Ni-P deposits with more than 9 wt.\% of phosphorus are entitled X-ray amorphous.

Owing to the variations of the pH value at the electrode-solution interface inherent to the Ni-P electroplating process, phosphorus content varies across Ni-P deposit thickness, hence its composition and microstructure are not homogeneous (Section \ref{Compositionally modulated coatings}). Alkaline solutions are more influenced by the pH variations compared to acidic ones, thus electrodeposits produced in this environment are often characterized by a lamellar structure \cite{Alleg_MicrostrAndMagnPropNiP_2016}. 

Ni-P electrodeposits can be grouped in three categories, alloys with: low (1-5 wt.\%), medium (5-8 wt.\%) and high (above 9 wt.\%) phosphorus content \cite{Daly_EChemNiPAlloyForm_2003}. Microstructure-wise, it is possible to fabricate polycrystalline, microcrystalline, nanocrystalline or fully amorphous Ni-P deposits through electroplating. 

Pillai et al. \cite{Pillai_EDNiPInDepthStPrepProp_2012a} found the structure of Ni-P alloy to undergo transition from crystalline to nanocrystalline and become amorphous at phosphorus contents above 9.14 wt.\%. Bredael and coworkers \cite{Bredael_JetCellRDElectrode_1993} fabricated Ni-P electrodeposits with phosphorus contents ranging from 0 to 18.0 wt.\% and a uniform composition profile across the samples. They observed that for phosphorus contents above 12 wt.\% the as-plated Ni-P coatings were amorphous, whereas below this threshold value, which was independent of the plating parameters, crystalline Ni-P coatings were obtained. In a subsequent work Bredael and colleagues \cite{Bredael_AmorpCrystEDNiP_1994} conducted a study with the goal to exactly identify the percentage of phosphorus at which the transition from crystalline to amorphous structure takes place. They found that the border between these two states can be found at phosphorus contents between 11.6 and 13.1 wt.\% P, with the structure being fully amorphous at P contents $\geq$13.1 wt.\%. Although there is no full consensus many authors agree that the phase transition of Ni-P happens over a wide range of wt.\% of P and not abruptly at a specific phosphorus content \cite{Pillai_EDNiPInDepthStPrepProp_2012a}. Lin et al. \cite{LinLeeChen_ElectrodepNiPSulfamBath_2006} found that not only phosphorus content influences the crystallographic structure of Ni-P electrodeposits. At the same phosphorus percentage substituting direct current deposition by pulse current plating can result in preserving crystalline structure even at phosphorus contents higher than a critical value. Contrary, phosphorus content at which transition from crystalline to amorphous structure occurs has been reported to decrease in the presence of certain additives, such as saccharin for example \cite{Daly_EChemNiPAlloyForm_2003}.

In general, discrepancies which exist in the literature related to the structural transitions of Ni-P electrodeposits can be attributed to the lack of appropriate techniques that would allow to make a proper differentiation between different crystallographic structures. Additionally, Ni-P alloys are frequently not straightforward amorphous or crystalline but they represent a mixture of several phases. This can be owing to the low solubility of phosphorus into the nickel lattice. Namely, often Ni-P deposits contain more of the alloying element (P) than what the f.c.c. Ni matrix can dissolve, thus the surplus must separate out resulting in regions that have a different concentration of phosphorus, different lattice parameters and different crystallite sizes. Vafaei-Makhsoos et al. \cite{Vafaei-Makhsoos_ElectrMicrCrystAmorphNiP_1978} found alloys with 3.8 and 6.7 wt.\% of P to be solid solutions composed of micrometer-sized grains. However, amorphous alloys (11.7 and 13 wt.\% P) were not homogeneous on a microscale and films contained microcrystalline regions. Ni-P electrodeposits displaying a structure that represents a mixture of solid Ni(P) solutions and amorphous or nanocrystalline phases have been reported \cite{Alleg_MicrostrAndMagnPropNiP_2016}.

Microstructure of Ni-P electrodeposits undergoes a transformation if they are exposed to a subsequent thermal treatment. This aspect is further elaborated in Section \ref{Thermal treatment}.

Suitable structure of the electroplated Ni-P alloy depends on the desired features of the fabricated deposit. Crystalline materials are traditionally interesting on account of their exceptional mechanical properties. Amorphous materials are very appealing owing to the isotropy of their properties and them lacking disadvantages characteristic for crystalline ones, namely the presence of crystal boundaries, lattice defects, segregation, etc. \cite{Zhaoheng_EDAmorphNiPAlloyCoat_1997}. Amorphous Ni-P electrodeposits are reported to possess better corrosion resistance when compared to crystalline deposits. However, transition to amorphous structure and consequent decrease of the grain size causes deterioration in mechanical properties which is termed an inverse Hall-Petch effect \cite{Schitz_AtomScSimul_1999}. Its onset is related to the increase of the volume fraction of the triple junctions  relative to the volume fraction of the grain boundaries which exerts a detrimental effect on coating's hardness \cite{Palumbo_TripLinDislEffect_1990}. Jeong et al. \cite{Jeong_EffofHeatTreatWearNiP_2003} produced Ni-P electrodeposits with grain size of less than \SI{10}{\nano\meter}, thereby achieving hardness values that fall into the range of the inverse Hall–Petch behaviour. Recently extensively reported are the results testifying even better mechanical properties of nanocrystalline materials (Figure \ref{fig:nanocryst_hardness}) when compared to both coarse-grained and amorphous ones \cite{Bicelli_ReviewNanostrAspMetED_2008,Low_MultifNanostrCoatED_2015,GurrappaBinder_EDNanostrCoatCharact_2008,AchantaCelis_NanocoatForTribAppl_2011}. The main feature, which makes nanocrystalline materials distinct from the other two mentioned groups, is the existence of a larger number of atoms disposed at interfaces, as grain boundaries and triple junctions. These interfaces are considered to be involved in the deformation mechanism occurring in the material. It is postulated that for nanocrystalline materials dislocations inside the grains hardly occur and that other plastic deformation mechanisms, such as grain boundary diffusion and sliding, grain rotation and a viscous flow of interfaces take place \cite{Apachitei_AutocatNionAl_2000}.
\subsection{Variables influencing phosphorus content}
Phosphorus content of Ni-P electrodeposits reportedly depends on many factors, dominantly on the concentration of the phosphorus donor in the bath, temperature, pH, current density, waveform in the case of pulse current deposition, agitation rate and the presence of additives. Thus, phosphorus quantity can be controlled by altering the electroplating bath composition and the employed deposition conditions. 

In the 1950s, Brenner \cite{Brenner_EDofQlloys_1963a} found that the amount of phosphorus incorporated in the Ni-P alloy increases with the increase of phosphorus acid content in the bath and with the decrease of the current density. However, too high H\textsubscript{3}PO\textsubscript{3} concentration leads to deterioration of the current efficiency owing to more cathodic charges being spent on the reduction of protons. Additionally, proton reduction results in the atomic and molecular forms of hydrogen, both of which can be incorporated in the deposit imposing inauspicious effects on its properties and increasing its internal stress \cite{Lin_StrEvolInterStrNiPED_2005}. Chang et al. \cite{Chang_EDNiPSulfamRelpHStr_2008} found that for the same content of H\textsubscript{3}PO\textsubscript{3}, phosphorous content can be increased by decreasing electrolytic bath pH value on account of the consequent increase of the concentration of non-dissociated phosphorous acid molecules in the electrolyte. This effect is however weakened at very low phosphorus acid concentrations on account of diffusion becoming the rate-controlling process.

Yuan and coworkers \cite{YuanSunYu_PrepOfAmorphNanocrNiP_2007} conducted a study on the preparation of amorphous-nanocrystalline Ni-P electrodeposits from a Brenner type plating bath. Their goal was to interrogate the key electroplating factors and their influence on the deposit's phosphorus content. Effects of: temperature, current density, pH, H\textsubscript{3}PO\textsubscript{3} concentration and agitation rate were investigated in the orthogonal experimental design study. Findings indicated that only pH value and H\textsubscript{3}PO\textsubscript{3} concentration and their interaction are the key variables affecting phosphorus content in the deposit. Figure \ref{fig:Amorph_Nanocr_NiP} shows contour plots for the constant phosphorus content versus Ni-P electroplating pH and H\textsubscript{3}PO\textsubscript{3} concentration.

Hu and Bai \cite{Hu_CompContrEPNiP_2001a} employed modified Watts Ni bath and similarly investigated the influence of the main electroplating variables, namely: temperature, current density, pH, NaH\textsubscript{2}PO\textsubscript{2}·H\textsubscript{2}O concentration and agitation rate on the phosphorus content in the fabricated Ni-P deposits. By using experimental strategies such as: fractional factorial design, path of steepest ascent and central composite design coupled with response surface methodology they came to the conclusion that the predominant factors affecting phosphorus content are temperature and current density of the electroplating with strong interactive effect between current density and pH. Figure \ref{fig:NiP_Comp_Contr} shows contour plots for the constant phosphorus content versus Ni-P electroplating temperature and current density.
\begin{center}
	\captionsetup{type=figure}
	\includegraphics[width=0.7\columnwidth]{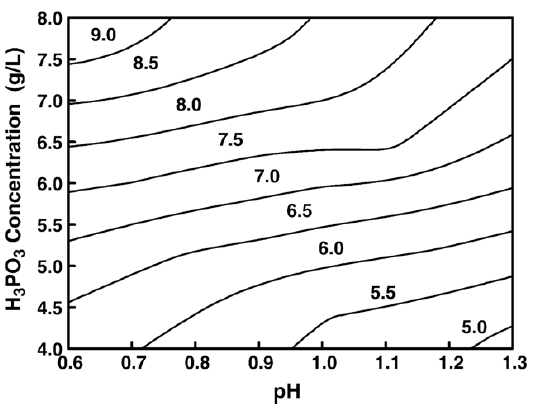}
	\captionof{figure}{Contour plots for the constant phosphorus content (P wt.\%) of Ni-P deposits against the electroplating pH (X\textsubscript{C}) and H\textsubscript{3}PO\textsubscript{3} concentration (X\textsubscript{D}). Reprinted with permission from \cite{YuanSunYu_PrepOfAmorphNanocrNiP_2007}. Copyright (2007) Elsevier.}
	\label{fig:Amorph_Nanocr_NiP}
\end{center}
\begin{center}
	\captionsetup{type=figure}
	\includegraphics[width=0.8\columnwidth]{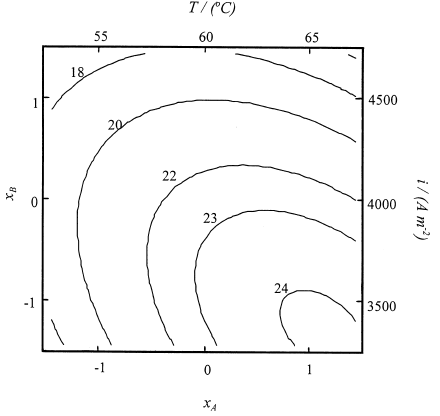}
	\captionof{figure}{Contour plots for the constant phosphorus content (P at.\%) of Ni-P deposits against the electroplating temperature (x\textsubscript{A}) and current density (x\textsubscript{B}). Reprinted with permission from \cite{Hu_CompContrEPNiP_2001a}. Copyright (2001) Elsevier.}
	\label{fig:NiP_Comp_Contr}
\end{center} 
\begin{center}
	\captionsetup{type=figure}
	\includegraphics[width=0.75\columnwidth]{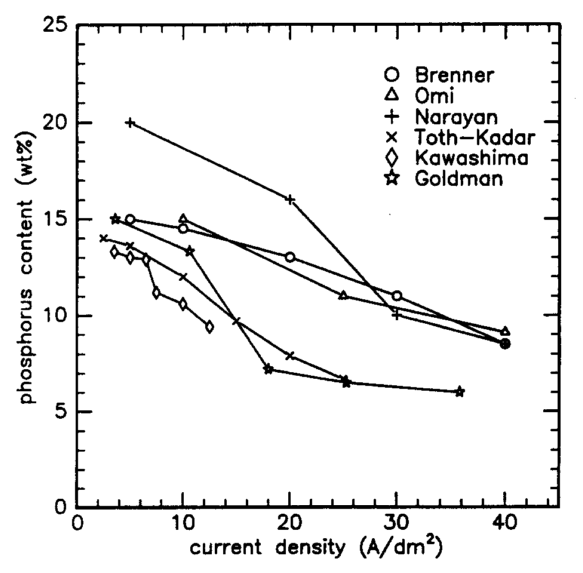}
	\captionof{figure}{Literature results on phosphorus content vs current density (up to \SI{40}{\ampere\per\square\deci\meter}). Reprinted with permission from \cite{Bredael_JetCellRDElectrode_1993}. Copyright (1993) Elsevier.}
	\label{fig:curr_dens_infl}
\end{center}

Recurring trend of gradual decrease of phosphorous content with the increase of the applied current density is observed in the work of many authors, however a large scatter between the data points of different authors can be perceived (Figure \ref{fig:curr_dens_infl}) \cite{Bredael_JetCellRDElectrode_1993}.

Bredael and coworkers \cite{Bredael_JetCellRDElectrode_1993} observed, in their study of Ni-P electrodeposition on a rotating-disc electrode, a steep transition from high to low phosphorus content with increasing local current density, while the literature data according to them do not show such a steep transition owing to the general approach where average current densities are used. In the bath where phosphorous acid is the only electrochemically active phosphorus species, dependence of the phosphorus content on the current density can be explained in terms of slow diffusivity of the large phosphorus acid ions compared with Ni\textsuperscript{+2} ions \cite{Bredael_JetCellRDElectrode_1993}. Pillai et al.\cite{Pillai_EDNiPInDepthStPrepProp_2012a} found that at higher H\textsubscript{3}PO\textsubscript{3} concentration in the bath ($\geq$ \SI{15}{\gram\per\liter}) phosphorus content in the coating is independent of current density. However, at low phosphorous acid amounts in the bath, the decreasing trend of phosphorus content in the coating with increasing current density was evident. Thus, they concluded that the observed scatter in the results presented by different authors, when it comes to the influence of current density on phosphorus content, can be mainly due to the different bath compositions used. In their study, an increase in the concentration of H\textsubscript{3}PO\textsubscript{4} resulted in no significant effect on the phosphorus content in the coating although the rate of deposition decreased continuously. An increase in the plating temperature (from \SI{50}{\degreeCelsius} to \SI{80}{\degreeCelsius}) resulted in the decrease of the amount of phosphorous incorporated in the coating and in the increase of the deposition rate. Naryan and Mungole \cite{Narayan_EDNiPalloy_1985} observed contrary to the previously mentioned study a slight increase of the phosphorus amount in the coating up to H\textsubscript{3}PO\textsubscript{4} concentrations of \SI{125}{\gram\per\liter} and a slight decrease with further increase in H\textsubscript{3}PO\textsubscript{4} concentration, all at the bath temperature of \SI{80}{\degreeCelsius}. At \SI{90}{\degreeCelsius} an increase in the H\textsubscript{3}PO\textsubscript{4} concentration generally produced a slight decrease in the phosphorus content in the coating. Naryan and Mungole additionally found that at low H\textsubscript{3}PO\textsubscript{3} concentrations, more phosphorus was deposited at \SI{90}{\degreeCelsius} than at \SI{80}{\degreeCelsius}, but at H\textsubscript{3}PO\textsubscript{3} contents in excess of \SI{25}{\gram\per\liter} more phosphorus was deposited at lower temperature. Sadeghi \cite{Sadeghi_MicrStrEvolStrMechNiComp_2016} employed hypophosphite as a phosphorous source and found that increasing its concentration and decreasing the current density (1-\SI{4}{\ampere\per\square\deci\meter}) both caused higher phosphorous contents in the fabricated Ni-P deposits. However, the content of phosphorus in the deposits electroplated from the baths containing very low phosphorus source amounts did not vary appreciably with the current density. Deposits obtained from the baths containing up to \SI{10}{\gram\per\liter} of NaH\textsubscript{2}PO\textsubscript{2} exhibited a decrease in the content of phosphorus as the current density was raised.

The increase in temperature has been observed to cause a decrease in phosphorus content in general. However, too low temperature leads to unsatisfactory current efficiency and plating rate, thus a compromise is needed. Temperatures from around \SI{50}{\degreeCelsius} to \SI{70}{\degreeCelsius} are generally adopted as acceptable for optimal feasibility of the Ni-P electroplating process. 

Phosphorus content in the deposit has been reported to be influenced by the presence of certain additives in the electroplating bath. Increase of the saccharin concentration was found to induce the decrease of phosphorus content \cite{Daly_EChemNiPAlloyForm_2003,Hansal_PCEDniPSiC_2013,A.Zoikis-Karathanasis2015}.

Influence of the pulse plating on Ni-P alloy structure and composition will be further elaborated in Section \ref{Pulse plating}.
\section{Properties of the electroplated Ni-P alloys}
\subsection{Mechanical, tribological and corrosion properties}
Alloying nickel with phosphorus via electroplating effectuates many improvements of its properties: increased hardness ($\gtrsim$500 HV as-plated Ni-P), wear and corrosion resistance, decreased coefficient of friction (0.4-0.7 as-plated Ni-P), but also transition towards paramagnetic features. Overall, properties and functional behaviour of Ni-P electrodeposits depend on their composition and microstructure. Ni-P coatings with low phosphorus content demonstrate high hardness and good wear resistance, while coatings with higher phosphorus content exhibit good corrosion resistance but poor mechanical properties owing to the transition towards amorphous structure. Amorphous Ni-P deposits are generally brittle and possess low ductility. At the same phosphorus content, microstructure of the deposit can be modified by altering the deposition conditions. Applying pulse plating instead of direct current plating can help to preserve Ni-P crystalline structure even at higher incorporated phosphorus quantities, thus good mechanical properties characteristic for low-phosphorus deposits can be maintained \cite{LinLeeChen_ElectrodepNiPSulfamBath_2006}.

Initially, microhardness of the alloy increases with phosphorus content augmentation at very low phosphorous amounts, while further enhancement of phosphorous quantity leads to hardness deterioration \cite{BerkhZahavi_EDPropNiPandComp_1996}. Figure \ref{fig:nanocryst_hardness} shows the evolution of the material's hardness with the change in its grain size. Nava et al. \cite{Nava_EffHeatTreatTribCorrNiP_2013} found that the wear volume of Ni-P electrodeposits is inversely proportional to the microhardness of the deposits. They perceived also that nanocrystalline Ni-P coatings, with phosphorus content of 2-8 \%, exhibit significantly higher wear resistance compared to coatings with higher phosphorus content. Pillai et al. \cite{Pillai_EDNiPInDepthStPrepProp_2012a} observed that Ni-P electrodeposits with phosphorus content in the range of 4-7 wt.\% exhibit good microhardness (7.74–8.57 GPa) and the microhardness of these alloys increases up to 12 GPa by annealing at \SI{400}{\degreeCelsius} during \SI{1}{\hour}. Yuan and colleagues \cite{YuanSunYu_PrepOfAmorphNanocrNiP_2007} reported that nanocrystalline composites with phosphorus content 5-9 wt.\% demonstrate better corrosion resistance than that of microcrystalline Ni-P deposits, and better abrasion resistance than that of amorphous Ni-P deposits. Therefore, under certain work conditions, when a compromise is needed in terms of obtaining both corrosion resistance and favourable mechanical properties, nanocrystalline deposits could present an optimal choice. 

Another way to ensure good mechanical and electrochemical properties is to design multilayered or graded coatings. For example, a duplex coating that consists of an inner layer having high phosphorus content and an outer layer with low phosphorus content presents a structure that combines advantageous features characteristic for each of the two compositions \cite{Luo_SynthDuplexNiPYSZNiP_2017,Wang_NovEDGradNiPReplHardCr_2006a,Wang_CorrResLubrBehNovGradNiPReplHardCr_2006}. Fabricating graded structures can also help to improve coating`s adhesion on the substrate owing to the gradual structure evolution across layer thickness and the lack of abrupt interfaces (Section \ref{Compositionally modulated coatings}). 

Adhesion of the coating on the substrate's surface is critical for all possible functional applications. Ni-P electrodeposits exhibit good adhesion on a range of widely differing substrates (brass, soft steel, copper, etc.). However, stainless steel for example can be problematic owing to the formation of a passive oxide layer which prevents Ni-P coating from sticking to its surface. Conventional way of addressing this issue is to apply on the substrate surface a Wood's or another nickel strike before Ni-P electroplating \cite{Dini_PlatOnDiffMetAlloys_1980}. Being highly acidic it dissolves the oxide and concurrently forms a thin layer of nickel on the stainless steel surface.

Published data offer diverse information regarding corrosion characteristics of Ni-P alloy, particularly about the nature of its anodic dissolution, ability to passivate and pitting susceptibility. Corrosion resistance of Ni-P coatings is much better than in the case of pure Ni \cite{BerkhZahavi_EDPropNiPandComp_1996}. In chloride containing and slightly acidic environments Ni-P electrodeposits demonstrate lower corrosion resistance when compared to neutral or slightly alkaline settings \cite{Bozzini_AnodBehNiPSnAcidClSol_2003}.
\begin{center}
	\captionsetup{type=figure}
	\includegraphics[width=0.75\columnwidth]{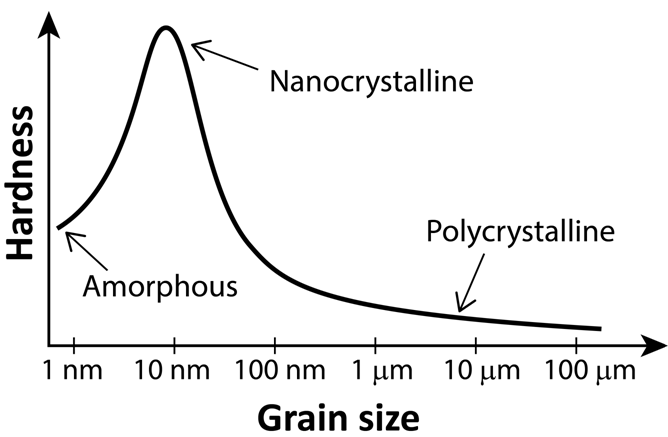}
	\captionof{figure}{Hardness of a material as a function of its grain size \cite{Low_MultifNanostrCoatED_2015}.}
	\label{fig:nanocryst_hardness}
\end{center} 

Amorphous and crystalline deposits have been found to exhibit different corrosion behaviour \cite{Daly_EChemNiPAlloyForm_2003}. Generally the weight loss of Ni-P deposits in corrosive solutions decreases with the increase of their phosphorus content \cite{Bai_CorrBehNiPBrine_2003}. Splinter and coworkers \cite{Splinter_XPSCHaractCorrFilmNiP_1996} reported the preferential dissolution of Ni on the surface of the Ni-P coating and the surface with the enriched P content after corrosion. They asserted that the formation of Ni\textsubscript{3}(PO\textsubscript{4})\textsubscript{2} film on the surface impedes the speed of corrosion. Additionally, their study demonstrated that the corrosion behaviour of nanocrystalline Ni-P alloys with lower phosphorus contents (1.4 wt.\% and 1.9 wt.\%) in 0.1M H\textsubscript{2}SO\textsubscript{4} approaches that of amorphous Ni-P electrodeposits (6.2 wt.\% P). X-ray photoelectron spectroscopy revealed that neither the nanocrystalline (1.9 wt.\% P, grain size 8.4 nm) nor the amorphous (6.2 wt.\% P) Ni-P alloys formed a passive layer in this environment. Diegle et al. \cite{Diegle_XPSPassivAmorphNi-20P_1988} reported that the addition of phosphorus improves the corrosion resistance of nickel owing to the reaction with water in which hypophosphite ions are formed. These ions prevent further dissolution of nickel through chemical passivation. Krolikovski and Butkiewicz \cite{Krolikowski_AnodBehNiPImpendSpec_1993} determined that the behaviour of Ni-P alloys is similar in neutral solutions under open circuit potential. However, under conditions of anodic polarization amorphous alloys exhibit dissolution suppression while in case of crystalline alloys intensive dissolution occurs.
\begin{figure*}[htb!]
	\centering    
	\includegraphics[height=85mm]{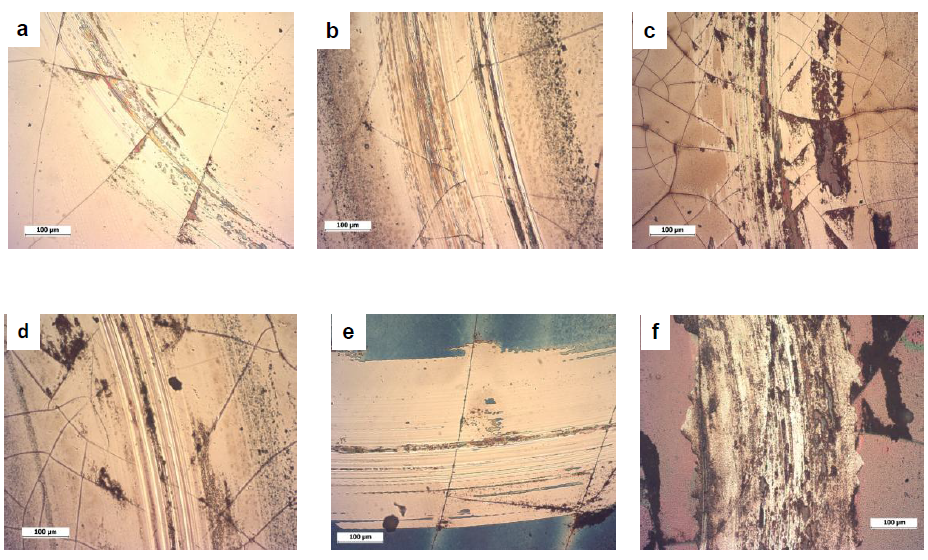}
	\caption{SEM images of wear track of Ni-P coatings, heat treated at different temperatures after sliding against AISI 8620 ball in air: a) without heat treatment, b) 200, c) 300, d) 400, e) 500, f) \SI{600}{\degreeCelsius}. Reprinted with permission from \cite{Nava_EffHeatTreatTribCorrNiP_2013}. Copyright (2013) ESG.}
	\label{fig:NiP_Wear}
\end{figure*}
\subsection{Internal stress}
Well controlled internal stress is a very important criteria for successful fabrication and use of protective Ni-P coatings. Compressive stresses are advantageous in mechanical structures under load, since they tend to inhibit crack formation and growth, they rise strength and hardness of the coating \cite{Sadeghi_MicrStrEvolStrMechNiComp_2016}. Factors influencing stress are many \cite{Baudrand_NiSulfamMystandPract_1996a}. Among them, the choice of the electroplating bath composition can very much aid in obtaining deposits with low values of this quantity. Baudrand \cite{Baudrand_NiSulfamMystandPract_1996a} fabricated electrodeposits from a sulfamate bath without the presence of additives with stress value of approximately 30 MPa, while the minimal value of stress for deposits obtained from a Watts bath was 180 MPa. 

The amount of phosphorus, hence crystallographic structure, of the Ni-P electrodeposit influences its stress value. Lin et al. \cite{Lin_StrEvolInterStrNiPED_2005} studied Ni-P electrodeposition from a sulfamate bath containing no additives. They observed that amorphous Ni-P deposits exhibit lower internal stresses compared to crystalline ones. The internal stress and the amount of adsorbed hydrogen exhibited a maximum at intermediate phosphorus contents suggesting that hydrogen incorporation and subsequent escape play an important role when it comes to the internal stress of the Ni-P electrodeposits. Nonetheless, high phosphorus content characteristic for amorphous Ni-P alloys comes with lower cathode current efficiency, thus conditions must be carefully optimized in order to secure optimal feasibility of the process along with desired properties of the obtained deposits. 

Certain organic additives are known to induce a reduction of the internal stress of the Ni-P deposits. A grain refiner saccharin is typically employed for this purpose, inducing compressive internal stress \cite{Daly_EChemNiPAlloyForm_2003,DiBari_EDNi_2011,Bicelli_ReviewNanostrAspMetED_2008,Bonino_ThermStabEDNiP_1997}. Nevertheless, additives containing sulfur can cause deterioration in the corrosion resistance of the fabricated coatings hence their application must be carefully optimized \cite{Osaka_SulfurInflOnCorrProp_1999}. 

Another possibility to reduce stress of a Ni-P electrodeposit is to apply a low temperature heat treatment that would help to desorb hydrogen present in the coating \cite{Daly_EChemNiPAlloyForm_2003}. 

Employing pulse current instead of direct current deposition is reported to be beneficial for stress reduction. Chen et al. \cite{Chen_IntStressContrNiPPC_2010} fabricated Ni-P deposits with high phosphorus contents and low internal stresses, ranging from tensile to compressive, with high current efficiency by using pulse current in a nickel sulfamate bath without the addition of any stress reducers. 
\subsection{Magnetic properties}
Nickel is a typical ferromagnetic material, while magnetic properties of the Ni-P alloy present a function of its composition and preparation technique \cite{Bakonyi_MagnPropEDMeltQLiqNiP_1993}. Magnetic moment and Curie temperature are found to decrease with the introduction of phosphorus into Ni matrix \cite{Albert_EffOfPonMagnOfNi_1967}. According to Weiss theory, nickel exhibits ferromagnetic properties owing to quantum-mechanical exchange forces which cause the spins of vicinal Ni atoms to be parallel \cite{Cullity_IntrMagnMat_2009}. The introduction of phosphorus into nickel matrix enlarges the separation between Ni atoms and with the increase of interatomic distance exchange forces decrease rapidly. Stated effect renders the transition from ferromagnetic to paramagnetic properties \cite{Dhanapal_EffofPonMagnPropNi-P_2015}. 

Bakonyi et al. \cite{Bakonyi_MagnPropEDMeltQLiqNiP_1993} performed a study of the magnetic properties of Ni\textsubscript{100-x}P\textsubscript{x} alloys fabricated by employing various techniques. For electrodeposited alloys the composition range studied was 11$\gtrsim$x$\lesssim$23. Throughout the whole concentration range magnetic inhomogeneities were observed. Alloys were found to exhibit paramagnetism at x$\gtrsim$17. Hu and Bai \cite{Hu_InfPContPhysChemPropNiP_2003} observed that the ferromagnetic property of Ni was transformed into paramagnetism at phosphorus content of 17 at.\% . They corroborated that the magnetic properties of Ni-P alloys are a function of their phosphorus content. In a subsequent study \cite{Bai_EffAnnealPhysChemNiP_2003}, they demonstrated that the paramagnetic Ni-P deposits become ferromagnetic after thermal treatment at \SI{400}{\degreeCelsius} owing to Ni and Ni\textsubscript{3}P phases separation. Knyazev and colleagues \cite{KnyazevFishgoitChernavskii_MagnPropOfEDAmorphNiP_2017} performed magnetization measurements and differential scanning calorimetry analysis of the Ni-P alloys obtained via electrodeposition. Their results indicated that the untreated alloys with phosphorus contents exceeding 12 at.\% were paramagnetic, owing to the lack of exchange interactions due to fluctuations in chemical composition and the formation of a network of phosphorus rich paramagnetic domains. Amorphous Ni-P alloys that were originally paramagnetic were rendered ferromagnetic through thermal treatment that also led to their devitrification. Dhanapal et al. \cite{Dhanapal_EffofPonMagnPropNi-P_2015} studied the influence of phosphorus content, employed duty cycle and current density on the magnetic properties of Ni-P alloy fabricated by pulse current deposition. They observed saturation magnetization dependence on the phosphorus content. At low phosphorus amounts saturation magnetization value of the obtained deposits was high. Increasing the duty cycle resulted in the increase of soft ferromagnetic nature of the Ni-P alloy and decrease of the coercivity and retentivity values.
\subsection{Catalytic activity}
Ni-P alloy is known to exhibit catalytic activity foremost in water splitting reaction. This presents a very important feature especially in the light of today's rapidly increasing energy demands and progressive exhaustion of fossil fuels resources. Catalysts (platinum and ruthenium) conventionally employed to reduce the large overpotentials for hydrogen and oxygen evolution reactions (HER and OER respectively) are fairly scarce and their use not feasible when it comes to incurred costs, thus suitable alternatives are in high demand. Nickel is a very popular electrode material owing to its reasonably high hydrogen generation activity, availability as well as its low cost \cite{Zeng_RecProgrAlkWatElectrHER_2010}. Although Ni as a catalyst does not perform as well as steel from the viewpoint of its electro-catalytic activity, it possess an excellent resistance to corrosion in hot concentrated alkaline solutions \cite{Safizadeh_ElectrkatDevForHErAlkEnv_2015}. Issues encountered when employing this metal as cathode material are its low catalytic activity or low resistance to intermittent electrolysis, however alloying nickel with other elements (P, Mo, etc.) can aid in mitigating these issues \cite{Safizadeh_ElectrkatDevForHErAlkEnv_2015,Ezaki_HydrOverPotTransMetAlloys_1993}. According to Paseka \cite{Paseka_HERNiP_2008}, the reason for the improvement of nickel activity in HER achieved through alloying it with phosphorus is the augmentation of the amount of amorphous phase surrounding the Ni crystals. Alloys containing amorphous phase which is able to dissolve large amounts of hydrogen possess high internal stress which contributes to their good electro-catalytic activity.

Hu and Bai \cite{Hu_OptOfHydrEvolvActNiP_2001} conducted a study in order to interrogate the effects of key electroplating variables on the hydrogen evolution activity of Ni-P deposits. By employing fractional factorial design, path of steepest ascent and central composite design, they found that the key factors influencing catalytic activity of Ni-P electrodeposits include: temperature, pH and NaH\textsubscript{2}PO\textsubscript{2}·H\textsubscript{2}O concentration and their interactions. The models they developed indicated that the alloy containing 7 at.\% of P should exhibit maximal electro-catalytic activity. In the subsequent study \cite{Hu_InfPContPhysChemPropNiP_2003}, they experimentally found deposit with 8 at.\% of P to be the best electrode material for HER. They attributed this deposit's highest specific activity to its largest true surface area and hence its maximum roughness. Wei et al. \cite{Wei_StdHydrEvReactNiPAmorphAlloy_2007} investigated catalytic activity of Ni-P in HER both experimentally and theoretically. Alloys that contain 10,8 at.\% P were found to perform the best. In order to elaborate the influence of phosphorus content in Ni-P amorphous alloys on their catalytic activity they employed density-functional theory and front molecular orbital theory. Obtained results indicated that alloys with phosphorus content anywhere from 9.1 at.\% to 14.3 at.\% should exhibit optimal activity for the whole HER. Paseka \cite{Paseka_HERNiP_2008} found that the alloy deposited at \SI{65}{\degreeCelsius} demonstrates worse catalytic activity than alloys deposited at temperatures $\leq$ \SI{53}{\degreeCelsius}. He asserted that this is owing to the presence of larger amounts of adsorbed hydrogen in alloys fabricated at lower temperatures, which contributes to their higher internal stresses. According to Paseka, catalytic activity also depends on the deposit's thickness and exhibits an increase as deposit grows. Thick deposits possess great catalytic activity owing to their high internal stress, however their mechanical instability presents a problem and thickness increase will be beneficial only up to a critical width of coating at which it will suffer a final failure. In the study by Li et al. \cite{Li_EffAmmonLowTempEDNiP_2003} it was determined that at low bath temperatures high NH\textsubscript{4}Cl content in combination with low NaH\textsubscript{2}PO\textsubscript{2} concentration can contribute to creating thick films (6-8 at.\% of phosphorus) with optimum catalytic activity even at high current densities. However, ammonia concentration must be chosen carefully because exceeding the ammonia concentration above the certain limit can negatively affect the cathode current efficiency, hence decrease the plating rate \cite{Dadvand_EDNiPfromCitrBath_2016}. Bai and Hu \cite{Bai_EffAnnealPhysChemNiP_2003} observed that for Ni-P electrodeposits with P content from 0 to 28 at.\% the ability of catalysing hydrogen evolution decreases with increasing annealing temperature, concluding that annealed deposits are not suitable for HER. 

Elias et al. \cite{Elias_EDNiPThinFilmsAlkwatSplitt_2016} studied the efficiency of electrodeposited Ni-P for both HER and OER in alkaline media. They found that the alloy thin films with 9.0 wt.\% of P and 4.2 wt.\% of P are the best electrode materials for HER and OER. Recently Tang and coworkers \cite{Tang_EDNiPHEROER_2015} reported a room-temperature electrodeposition of Ni-P nanoparticle film on Ni foam which acts as a bifunctonal water-splitting catalyst in strongly alkaline media with very small overpotentials for HER and OER: \SI{80}{\milli\volt} and \SI{309}{\milli\volt}, respectively. 

As an alternative to alkaline media in which water electrolysis systems are simple but demonstrate low efficiency and high energy consumption, Lu et al. \cite{Lu_ElectrokatPropNiAloyysHERAcid_2003} interrogated electrocatalytic activity of Ni based alloys in acidic media. They observed a decrease of catalytic activity with the increase of phosphorus content which they attributed to moving away from the optimal electronic configuration with phosphorus incorporation but additionally to the decrease of the number of grain boundaries which present active sights for HER. Contrary, authors of \cite{Kucernak_NiPEffectOfPcontHERAcid_2014} claimed that increasing the phosphorus content, even past contents that can be achieved through conventional electrodeposition would impart exceptional catalytic properties to these systems for applications in acidic media and hence very high phosphorus contents ought to be beneficial. 
\section{Thermal treatment of electroplated Ni-P alloy}
\label{Thermal treatment}
With subsequent heat treatment the hardness of Ni-P electrodeposits increases substantially. This phenomenon is caused by the precipitation of the crystalline phases. Most studies report detecting a mixture of Ni\textsubscript{3}P and f.c.c. nickel as a final product of the thermal treatment \cite{Nava_EffHeatTreatTribCorrNiP_2013}. 

The annealing temperature under which crystallization occurs depends on the phosphorous content in the deposit as well as on the heating rate \cite{Keong_CrystKinPhTransfNiPHighP_2002}. Phase transformation temperatures of the deposits increase with increasing heating rate and decreasing phosphorus content \cite{Keong_CrystKinPhTransfNiPHighP_2002,Bai_EffAnnealPhysChemNiP_2003}. Continued heating at temperatures higher than the transition temperature leads to a decrease of hardness owing to the subsequent recrystallization and grain coarsening \cite{BerkhZahavi_EDPropNiPandComp_1996,Keong_HardEvolELNiP_2003,Apachitei_EffHeatTrniPandNiPSiC_2002}. The degree of loss of hardness is higher for deposits having lower phosphorus content on account of more pronounced grain growth and coarsening of Ni phase compared to Ni\textsubscript{3}P phase \cite{Keong_HardEvolELNiP_2003}. Transition from amorphous to crystalline state is followed by a thermal contraction phenomena due to higher density of crystalline structure when compared to amorphous one, the amplitude of contraction being higher at higher phosphorus contents \cite{Bonino_ThermStabEDNiP_1997}.

Nava et al. \cite{Nava_EffHeatTreatTribCorrNiP_2013} fabricated Ni-P electrodeposits containing 10.6 at.\% of P which exhibited maximum hardness (990 HV) after thermal treatment at \SI{500}{\degreeCelsius} (comparable to the hardness of hard Cr coatings $\sim$1000 HV). Obtained deposits demonstrated also the lowest wear rate as indicated by the SEM images of their wear track patterns that exhibited the narrowest width and shallowest plough lines (Figure \ref{fig:NiP_Wear}). A linear relationship was detected between the hardness and the wear resistance of the heat treated Ni-P alloy coatings. Corrosion resistance of the deposits deteriorated upon annealing owing to the formation of cracked structure in the thermally treated coatings, which promoted localized corrosion. Bai and Hu \cite{Bai_EffAnnealPhysChemNiP_2003} found crystallization for Ni-P deposits containing $\leq$ 24 at.\% of P to occur at \SI{400}{\degreeCelsius}, while deposits with 28 at.\% demonstrated a phase transformation at \SI{200}{\degreeCelsius}. Additionally, phosphorus content in the deposits decreased with increasing the annealing temperature owing to the replacement of phosphorus by the oxygen from the air. 

Habazaki et al. \cite{Habazaki1_EffStrRelaxOnCorrBehNiP_1991} conducted a study in order to interrogate the effect of annealing on the microstructure and the corrosion behaviour of the electrodeposited amorphous Ni-P alloys. Obtained results indicated that Ni-P alloys with 19.2 at.\% phosphorus crystallize directly to f.c.c. Ni and Ni\textsubscript{3}P phases. High phosphorus alloys ($\geq$ 24.6 at.\% P) were first crystallized to a metastable single phase and then decomposed to Ni\textsubscript{3}P. Crystallization of alloys with the intermediate P content (19.2 at.\% < P < 24.6 at.\%) resulted first in a mixture of f.c.c. Ni, Ni\textsubscript{3}P and the metastable phase. The precipitation of f.c.c. Ni in the amorphous phase occurred for Ni-16.7P alloy before complete crystallization. Annealing decreased corrosion resistance for alloys containing $\leq$22.7 at.\% P owing to the formation of phosphorus deficient f.c.c. Ni phase. Keong et al. \cite{Keong_CrystKinPhTransfNiPHighP_2002} similarly reported that amorphous coatings with high phosphorus content follow a sequence of transformations during annealing and form metastable phases, such as Ni\textsubscript{2}P and Ni\textsubscript{12}P\textsubscript{5} before forming stable Ni\textsubscript{3}P and f.c.c. Ni phases. Zoikis et al. \cite{Zoikis-Karathanasis_PCEDNiPSiC_2010} observed crystallization of the amorphous Ni-P electrodeposit into Ni and Ni\textsubscript{3}P phases after annealing at $\sim$ \SI{400}{\degreeCelsius}. The presence of Ni\textsubscript{2}P phase after thermal treatment above \SI{330}{\degreeCelsius} was detected. Ni\textsubscript{2}P phase formation was also reported in \cite{SPYRELLIS_NiNi-PMatrixComp_2009}.  

Jeong et al. \cite{Jeong_RelHardAbrWearEDNiP_2003} reported the increase of hardness and abrasive wear resistance after heat treatment for nanocrystalline Ni-P coatings, with Taber abrasive wear resistance being linearly proportional to the hardness of the coatings. Heat treatment causes also the elastic modulus of the deposit to increase significantly according to the authors of \cite{Pillai_EDNiPInDepthStPrepProp_2012a}. When maximum hardness is achieved upon annealing fracture toughness exhibits the lowest value \cite{Yonezu_ContFrMechNiPStSubstr_2013}. 

Chang et al. \cite{Chang_StrMechNiPNanocrGr_2007} argued that significant strengthening by annealing for electrodeposited Ni-P alloys with low phosphorus content and nanocrystalline grains is not induced by the precipitation of Ni\textsubscript{3}P phase. The increase in hardness upon annealing is according to them a result of grain boundaries relaxation, phosphorus segregation, reduction of interior defects with the possible contribution from the increase of density owing to degassing of hydrogen.

Opting for subsequent Ni-P alloy thermal treatment and accordingly choosing the suitable heat treatment temperature, finally depend very much on the employed substrate nature and the intended process complexity.
\section{Application of pulse plating for Ni-P alloy fabrication}
\label{Pulse plating}
Constant direct current is the most commonly applied regime in which metallic coatings are electrodeposited. However, in recent years the use of pulsed, alternating currents is exhibiting an increase. This is owing to number of findings that demonstrate that pulse current (PC) exerts a beneficial influence on the structure and properties of the fabricated deposits. Among others, comprehensive reviews on pulse and pulse reversed plating are composed by Devaraj and Seshadri \cite{Devaraj_PP_1990} and by Chandrasekar and Pushpavanam \cite{Chandrasekar_PPandPRPConcAdvApp_2008}.

Development of modern electronics has granted a great flexibility in programming of the applied modulated current waveforms, with trains of pulses that can be programmed to have very complex sequences and forms. 
\begin{center}
	\captionsetup{type=figure}
	\includegraphics[width=0.65\columnwidth]{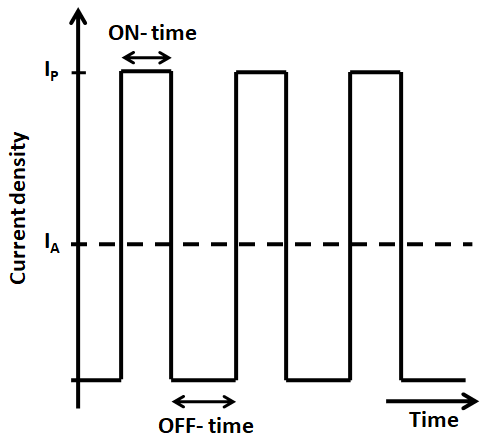}
	\captionof{figure}{Typical pulse-current waveform.}
	\label{fig:PC_waveform}
\end{center}

Rectanglular waves are the easiest waveforms to produce \cite{Pearson_FactFictPP_1991}. They have been demonstrated to produce a higher nucleation rate of the grains compared to triangular waveform \cite{HuChan_TriangWaveFNiSiC_2005}. Typical waveforms employed in pulse plating include: cathodic pulse followed by a period without current (or an anodic pulse), direct current (DC) with superimposed modulations, duplex pulse, pulse-on-pulse, cathodic pulses followed by anodic pulses (pulse reverse current-PRC), superimposing periodic reverse on high frequency pulse, modified sine-wave pulses and square-wave pulses \cite{Chandrasekar_PPandPRPConcAdvApp_2008}. 

During electroplating, a negatively charged layer is formed around the cathode as the process advances. When performing DC deposition, this layer charges to a definite thickness and prevents ions from reaching the substrate. In a simple form of PC electrodeposition, where cathodic pulse is alternatively switched on and off, when the output is turned off this layer discharges which allows easier transport of ions from the bulk of the solution \cite{Chandrasekar_PPandPRPConcAdvApp_2008}. Additionally, during plating high current density areas in the bath become more depleted of ions compared to low current density areas. During time when the current is off ions migrate to the depleted areas and when subsequent current pulse occurs more evenly distributed ions are available for deposition onto the part \cite{Chandrasekar_PPandPRPConcAdvApp_2008}. In the absence of current, small grains are recrystallized owing to their higher surface energy which makes them less thermodynamically stable than large grains, hydrogen is also desorbed decreasing the internal stress of the obtained deposits. In general, PC plating results in finer grain deposits exhibiting improved properties, including hardness, roughness, porosity, wear resistance, etc. Pulse plating can reduce additive requirements substantially \cite{Chandrasekar_PPandPRPConcAdvApp_2008,Devaraj_PP_1990}. 

Pulsed current results in metal deposition at the same rate as direct current provided that the average pulse current is equal to the mean direct current of DC electrodeposition \cite{Low_MultifNanostrCoatED_2015,Chandrasekar_PPandPRPConcAdvApp_2008}. 

In PC plating, choice of the applied waveform is critical and care must be taken to appropriately optimize all parameters (peak current, duty cycle, frequency, pulse shape, etc.). In his work Pearson \cite{Pearson_FactFictPP_1991} explores the benefits and the limits of the PC electrodeposition technique. He asserts that very low duty cycles are not feasible, because in order to produce the same average deposition rate as for DC, as duty cycle is reduced the pulse peak current needs to be increased. In practical applications, too high peak current densities are seldom viable due to limitations of rectifiers capacities. On the other hand, as duty cycle is increased the process begins to approach direct current deposition, thus a compromise must be made. When it comes to the frequency, practical maximum frequency which can be applied is limited by the capacitance of the double layer at the interface between the plating electrolyte and the article being plated. If the frequency is very high, the double layer does not have enough time to fully charge during the pulse and the process begins to resemble DC deposition \cite{Pearson_FactFictPP_1991}. Maximum useful frequency is around \SI{500}{\hertz} for most applications. However, higher frequencies can be used where very high peak current densities are employed because the double layer charge and discharge times become shorter as the peak current density is increased. 

In pulse reversed current technique (PRC) anodic pulse is introduced into the plating cycle. PRC has the same effect of replenishing the diffusion layer as PC does. It results in dissolution of the protrusions on the metal surface ensuring a more uniform deposition through elimination of the discrepancies between high and low current density areas and increases coating thickness uniformity \cite{Chandrasekar_PPandPRPConcAdvApp_2008}. PRC plated amorphous Ni-P deposits are reported to exhibit better ductility owing to the absence of voids and to consist of layers with different amorphous structures \cite{Zeller_PRCNiP_1991}.

It has been reported that the application of PC deposition results in the increase in limiting current density \cite{Chandrasekar_PPandPRPConcAdvApp_2008}. However according to Pearson \cite{Pearson_FactFictPP_1991} total thickness of the diffusion layer is equivalent to that obtained when plating in DC regime, owing to this the use of PC has very little effect on the limiting current density.

In Ni electrodeposition, pulse plating is extensively employed and studied \cite{Kollia_NuiPlatingPCTexandMicrstrMod_1990,Kollia_MicrohAndRoughNiEDPRC_1993,Kollia_InflPCOrientSurfMorphniED_1991}. Pavlatou et al. \cite{Pavlatou_HardEffIndSiCIncNiMatr_2006} investigated the use of pulse current in order to interrupt the columnar growth of the nickel grains and to produce more compact and thus more corrosion resistant coatings. 

In Ni-P electroplating, pulse plating also possesses several advantages over DC plating. Lin and coworkers \cite{LinLeeChen_ElectrodepNiPSulfamBath_2006} studied Ni-P alloy deposition in a PC regime from a sulfamate bath. They established that compared with DC plating, current efficiency associated with high phosphorus content deposits can be improved by applying pulse current having low duty cycle, high frequency, and proper peak current density. It was demonstrated that after applying the same total charge associated with DC and PC waveforms of different duty cycles, a more uniform concentration profile is maintained for PC than for DC deposition, particularly with the PC having small duty cycles. By employing 0.1 duty cycle with frequencies exceeding 100 Hz deposits with 14 wt.\% of P were plated with an efficiency of around 80\%. Low duty cycles and high frequencies are beneficial in terms of maintaining more stable surface proton concentration distribution inhibiting alloy composition modulations induced by pH value variations. Additionally, unlike DC-plated Ni-P deposits that might become amorphous when their phosphorus content exceeds a critical value, pulse-plated deposit with 14 wt.\% phosphorus still consisted of equiaxed crystalline grains. This is in agreement with the study done by Chen et al. \cite{Chen_IntStressContrNiPPC_2010} who concluded that a reduction in duty cycle from 0.5 to 0.1 simultaneously increases phosphorus content and grain size of the deposits. They found that Ni-P deposits with high P content can be achieved with high current efficiencies by employing pulse plating and that the obtained deposits exhibit lower internal stress than the DC-plated coatings. Pulse-plated deposits were also consistently harder than the DC-plated ones.
\begin{center}
	\captionsetup{type=figure}
	\includegraphics[width=0.95\columnwidth]{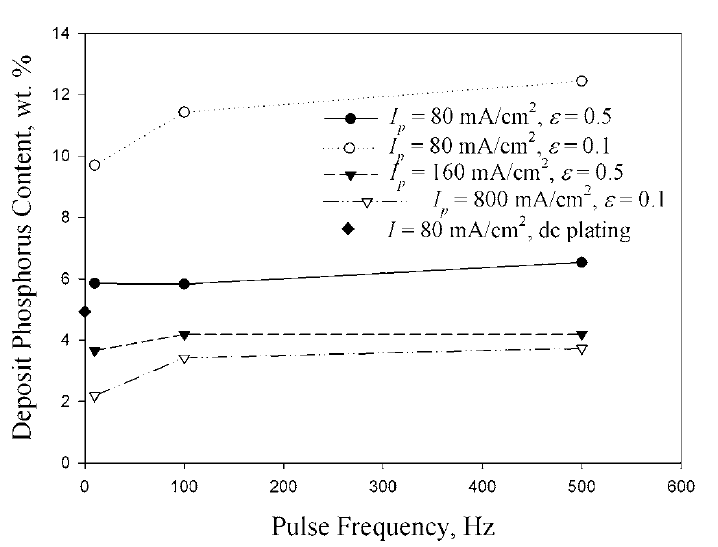}
	\captionof{figure}{Deposit phosphorus content as a function of the peak current density, pulse frequency, and duty cycle. Reprinted with permission from \cite{LinLeeChen_ElectrodepNiPSulfamBath_2006}. Copyright (2006) Electrochemical Society, Inc.}
	\label{fig:NiP_sulfam_Impr_CE_1}
\end{center}

\begin{center}
	\captionsetup{type=figure}
	\includegraphics[width=0.95\columnwidth]{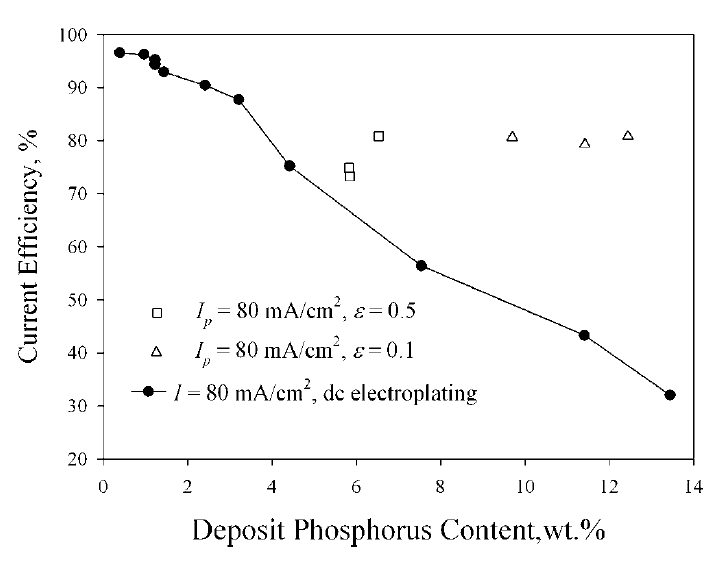}
	\captionof{figure}{Dependence of current efficiency on the deposit phosphorus content. Reprinted with permission from \cite{LinLeeChen_ElectrodepNiPSulfamBath_2006}. Copyright (2006) Electrochemical Society, Inc.}
	\label{fig:NiP_sulfam_Impr_CE_2}
\end{center}
\section{Compositionally modulated Ni-P electrodeposits}
\label{Compositionally modulated coatings}
In DC plating ways of improving deposit properties include microstructure manipulation by varying deposition conditions or alloying in which case a two-component deposit is formulated by employing an appropriate bath formulation. However, technological development imposes the demand for coatings that combine favourable properties of different metals, a task unachievable solely by employing individual components neither their resultant alloys. In this case compositionally modulated (functionally graded or multi-layered) coatings present them selves as a remarkable opportunity owing to the possibility of engineering coatings fitting a wide range of requirements in terms of functional properties.

In nickel electroplating, production of multilayer coatings consisting of films of bright and semi bright nickel (60-75\% of the total nickel thickness) is traditionally employed to improve deposits' corrosion resistance. Preferential corrosion in the electrochemically more active upper bright nickel layer and propagation of the corrosion in the lamellar direction protects the columnar lower layer of semi-bright nickel through retarded pitting attack penetration \cite{NiPlatingHandbook_2014}.

In the number of applications including the use of coatings for thermal, wear or corrosion protection and microelectronics, the mismatch in properties at the interface between the coating and the substrate can cause stress concentration that could result in the failure of the interface. Employing functionally graded materials (FGMs) \cite{Zhao_DevAndApplFunctGradMat_2012,Udupa_FunctGrdMatOverv_2014,Niino_RecDevStFGM_1990} which are characterized by a position-dependent chemical composition, microstructure or atomic order \cite{Wang_CorrResLubrBehNovGradNiPReplHardCr_2006} and by a subtle gradient of their properties along the coating thickness is reported to help in reducing the stress usually incurred due to the abrupt composition change when going from substrate toward deposit and decreasing the danger of its delamination.

A non-uniform, graded \cite{Wang_NovEDGradNiPReplHardCr_2006a} or layered distribution \cite{Gamburg_EDAlloysCM_2001,Roy_CMMMs_2009} relative to the phosphorus content along the coating thickness or a grain size gradient \cite{Qin_NovelEDNanostrNiGrainSizeGrDistr_2008,Gu_LayerNanostrNiModHard_2007} are deemed to be beneficial also for imparting favourable properties to Ni-P deposits. 

Phosphorus content in Ni-P electrodeposits inherently varies with layer thickness. pH at the electrode-solution interface rises concomitantly with the discharge of hydrogen. After escape of a cohort of hydrogen bubbles from the cathode resultant enhanced convection increases the interfacial proton concentration. Interfacial pH variations cause the variations in the deposit phosphorus content as electroplating continues \cite{Lin_StrEvolInterStrNiPED_2005}. Crousier et al. \cite{Crousier_CyclVoltStNiP_1993,Crousier_EDniPAmorphMultilStr_1994} confirmed that Ni-P electrodeposits consist alternately of layers having varying phosphorus contents. Sadeghi \cite{Sadeghi_MicrStrEvolStrMechNiComp_2016} observed that the formation of layers with different phosphorus amounts occurs more readily for alloys with low or medium phosphorus content and at higher current densities. A homogeneous deposit can be fabricated and stratification avoided by enhancing mass transport to the cathode surface or by employing pulse plating \cite{Lin_StrEvolInterStrNiPED_2005}. Additionally, oscillations in pH value are damped out as coating grows in thickness and electroplating continues \cite{Daly_EChemNiPAlloyForm_2003}. However variations in phosphorus content can be sometimes intentionally induced in order to produce composition modulation in fabricated Ni-P electrodeposits. With careful optimization this approach can result in a significant improvement of features when compared to conventional homogeneous Ni-P electrodeposits. Layered crystalline/amorphous Ni-P alloys are for example reported to exhibit higher tensile strength, moduli of elasticity and improved corrosion resistance \cite{Crousier_CyclVoltStNiP_1993}.   

It is known that Ni-P deposits with low phosphorus content exhibit high hardness and good wear resistance but poor corrosion resistance, conversely high phosphorus coatings exhibit good corrosion resistance but poor mechanical properties. Developing multilayer coatings can be an effective way to obtain deposits characterized by both optimal mechanical and electrochemical properties. For example, a duplex coating with an outer layer having a low phosphorus content and an inner layer with high phosphorus content is a good way to ensure corrosion stability in the contact with surrounding environment and in the same time favourable mechanical properties \cite{Luo_SynthDuplexNiPYSZNiP_2017}. 

Techniques in order to produce layered Ni-P electrodeposits include either using two baths with different phosphorus source contents or a single electroplating bath in which case electrochemical methods are employed in order to achieve composition modulation (changing cathodic current or potential). 

In the dual-bath electrodeposition technique, the item to be electroplated is moved between two plating baths of arbitrary composition and a layer is plated from each electrolyte in cycles. First multilayer deposition by employing dual bath technique was reported by Blum for Ni(\SI{24}{\micro\meter})/Cu(\SI{24}{\micro\meter}) as early as 1921 \cite{Blum_DualBath_1921}. This approach is mechanically complex compared to the single-bath method and carries a risk of contamination during the substrate transfer \cite{ROSS_EDtechniqMiltLay_1993}. Historically, first realization of single-bath electrodeposition was reported by Brenner and Pommer in 1948 who produced multilayered coatings CuBi($\sim$\si{\micro\meter})/BiCu($\sim$\si{\micro\meter}) by alternately switching the deposition current between low and high value \cite{Brenner_SinglBath_1948}. Single bath method is efficient, versatile and technologically simple. A review of both dual-bath and single-bath electrodeposition methods is compiled by Ross \cite{Ross_EDMultilayThinFilms_1994}.
\begin{figure*}[htb!]
	\centering    
	\includegraphics[height=60mm]{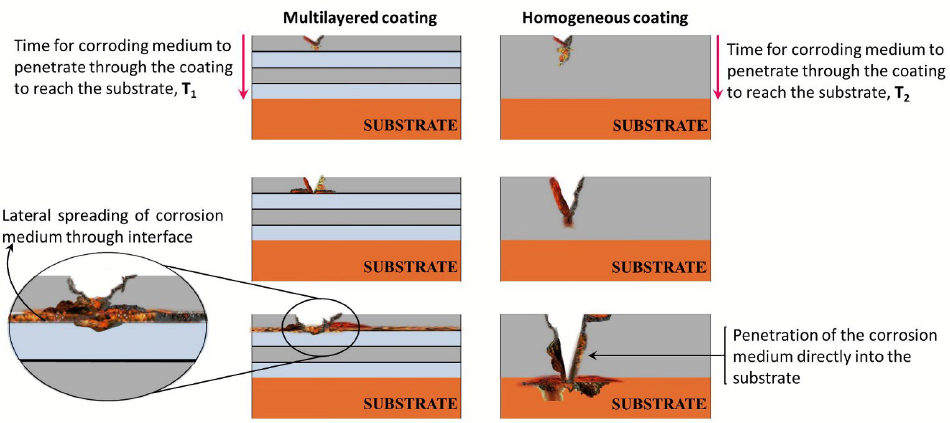}
	\caption{Diagram showing the mechanism of increased corrosion protection in the case of multilayer Ni-P coating (left), compared to monolayer Ni-P coating (right). It demonstrates that the time required for the corroding medium to reach the substrate by penetrating through the multilayer coating (T\textsubscript{1}) is much greater than that for the monolayer coating (T\textsubscript{2}). Reprinted with permission from \cite{Elias_DevNanoLamMultiLayNiPCorrProt_2016}. Copyright (2016) Royal Society of Chemistry.}
	\label{fig:Nanolam_NiP}
\end{figure*}

When employing single bath deposition method pulsed current is a useful tool to achieve the desired composition modulation. Early attempt with two-pulse plating was reported by Girard \cite{Girard_EDThinMagnPermal_1967} for the electrodeposition of permalloy films with a composition as close as possible to the zero-magnetostriction (Ni\textsubscript{81}Fe\textsubscript{19}) ensuring the smallest magnetic coercive force.

In the realm of compositionally modulated Ni-P electrodeposits, Goldman and coworkers \cite{Goldman_ShortvwCompMod_1986} demonstrated a method to fabricate Ni/Ni-P films, with wavelengths between 2.1 and 4.0 nm and an average phosphorus content around 12 at.\%, by alternating electrodeposition in two baths of different composition. Ross et al. \cite{ROSS_EDtechniqMiltLay_1993} revealed a dual bath electrodeposition technique for the production of thin-film metal multilayers in which substrate was suspended above nozzles of electrolyte and rotated by a motor. Specific steps were taken, including washing and drying the substrate with N\textsubscript{2}, in order to mitigate problems related to cross-contamination between electrolytes. Multilayered films of Ni/NiP\textsubscript{x}, NiP\textsubscript{x}/NiP\textsubscript{y}, Cu/Ni, and Co/NiP\textsubscript{x} were fabricated with a range of repeated lengths. Ni/NiP\textsubscript{x} and NiP\textsubscript{x}/NiP\textsubscript{y} multilayers exhibited the highest quality, with repeat lengths as low as \SI{19}{\angstrom} and up to three orders of reflection in low angle X-ray scans. NiP\textsubscript{x}/NiP\textsubscript{y} multilayers were fabricated over a wide range of compositions and with crystalline or amorphous structure. Problems were encountered with other compositions, in case of Cu/Ni and Cu/Co systems due to galvanic coupling as well as contamination and in case of Co/Ni-P owing to non-uniformity of Co nucleation. 

Wang et al. \cite{Wang_NovEDGradNiPReplHardCr_2006a} described electrodeposition and investigated properties of Ni-P deposits having a varying phosphorus content in the direction of the coating thickness. Single bath method was applied and composition gradient was achieved through varying current density (5-\SI{30}{\ampere\per\square\deci\meter}). The wear resistance of the fabricated Ni-P electrodeposits was approximately two times grater than that of the ungraded Ni-P deposits. Beneficial wear properties of the obtained deposits were attributed to the inhibition of formation and propagation of through-thickness cracks during the wear process owing to their graded structure. Heat treated coatings exhibited low friction coefficient and hardness that was close to the one of hard Cr coatings. In another study, Wang and colleagues \cite{Wang_CorrResLubrBehNovGradNiPReplHardCr_2006} examined corrosion resistance of the developed Ni-P gradient deposits and their tribological behaviour under the oil-lubricated conditions. Deposits heat treated at \SI{400}{\degreeCelsius} exhibited two orders of magnitude better corrosion resistance than hard Cr coatings (Figure \ref{fig:Corr_NiP_FGD}). The best corrosion resistance was found for deposits heat-treated at \SI{200}{\degreeCelsius}, which was attributed to the preserved amorphous structure and the stress relaxation at this temperature. Heat-treated coatings exhibited relatively higher wear rate and friction coefficients than hard Cr deposits under oil-lubricated wear conditions.
\begin{center}
	\captionsetup{type=figure}
	\includegraphics[width=0.85\columnwidth]{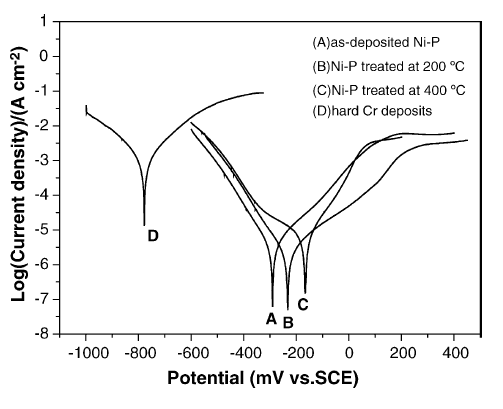}
	\captionof{figure}{Potentiodynamic polarization curves obtained for Ni-P graded deposits and hard Cr coatings, measured in 10 wt.\% HCl solution. Reprinted with permission from \cite{Wang_CorrResLubrBehNovGradNiPReplHardCr_2006}. Copyright (2006) Elsevier.}
	\label{fig:Corr_NiP_FGD}
\end{center}

Elias et al. \cite{Elias_DevNanoLamMultiLayNiPCorrProt_2016} fabricated multilayer Ni-P alloy coatings for better corrosion protection of mild steel by employing cyclic modulation of the cathode current density. Achieved improvement in the corrosion protection of fabricated multilayer Ni-P alloy coatings was attributed to the large number of interfaces between layers of alloys having different composition and phase structures, at which corrosion propagates laterally until the interface breaks down (Figure \ref{fig:Nanolam_NiP}). This mechanism systematically slows down corrosion and improves the coatings stability. Corrosion protection efficiency of multilayer coatings was found to increase with the number of layers, however only up to a certain point (300 layers). At very high layer numbers corrosion resistance deteriorated owing to the lack of distinct interfaces between individual films (approaching a monolayer structure). However, not only the number of layers but also their composition determined the performance of fabricated deposits. 

Multilayers of Ni-P with other metals or alloys can be engineered also in order to ensure optimal functional performance in demanding environments.

Improved corrosion properties of multilayers of Ni-P/Zn-Ni, when compared to pure Ni-P and Zn-Ni alloys, was detected by Liu and colleagues \cite{Liu_ZnNiP_2013}. According to them, this was due to the corrosion of the sacrificial sublayers of Zn-Ni which extends in the direction parallel to the substrate surface. Continued corrosion generation through the subsequent sublayers overall slows down the corrosion reaching the substrate and an eventual material brake down. Bahadormanesh and Ghorbani \cite{Bahadormanesh_ZnNiP_2017} recently devised a single bath deposition method for electrodeposition of Ni-P/Zn-Ni compositionally modulated multilayer coatings. At low current densities Ni–P was deposited, while at higher current densities Zn–Ni alloy containing 3.2 wt.\% P was obtained. It was observed that the Ni–P/Zn–Ni compositionally modulated multilayer coatings were sacrificial to the steel substrate.

Bozzini and colleagues \cite{Bozzini_AnodBehNiPSnAcidClSol_2003} fabricated Ni-P and Sn multilayer amorphous deposits (layer thickness \SIrange{0.1}{0.5}{\micro\meter}) by employing a dual bath electrodeposition technique in order to explore the possibility of improving the Ni-P deposits passivation behaviour. They investigated anodic behaviour of obtained coatings in acidic chloride solution. Current densities at the passivation plateau, were in the range of \SIrange{2}{5}{\micro\ampere\per\square\centi\metre} for Ni-P/Sn multi layers, in comparison to a passivation current density of \SI{45}{\micro\ampere\per\square\centi\metre} for Ni-P coatings with similar phosphorus content. They found that the present interfaces improve considerably the passivation behaviour provided that the incorporated Sn layers are thin enough. 
\section{Conclusion}
Ni-P protective coatings can be electrodeposited with a wide range of crystallographic structures. Deposits extending from fully crystalline to amorphous ones can be easily fabricated. Properties of the obtained coatings very much depend on the deposition conditions and their phosphorous content. Applied post treatment, such as alloy heating, can bring significant improvement of its overall mechanical, tribological and electrochemical properties. Designing unconventional structures, such as graded or multilayered ones, can also with proper optimization give rise to substantial deposits' characteristics amelioration.

However, even though technology of Ni-P electroplating is quite mature, there are still many unknowns and numerous issues still remain to be addressed. Phosphorus incorporation mechanism is still a matter of great disagreement. Factors influencing phosphorus content of the deposit are many and with the plethora of variables characterizing Ni-P electroplating process and poor definition of certain process parameters it is not easy to establish clearly the key influencers, manner and the extent of their impact. Up-scaling of the electroplating procedure induces new process variables and more attention in research needs to be bestowed on feasibility study of method transferral to larger scales and its robustness. Current density distribution is a parameter that is often poorly defined. Ageing of the electrolytic baths is not addressed in many of the research works. Additionally, distinguishing between different Ni-P microstructures and establishing a point of transition from one state to another is still a matter of some difficulty. 
\subsection*{Acknowledgements}
This work was supported by the European Union’s Horizon2020 research and innovation programme SOLUTION, under grant agreement No. 721642.
\begin{center}
	\captionsetup{type=figure}
	\includegraphics[width=0.95\columnwidth]{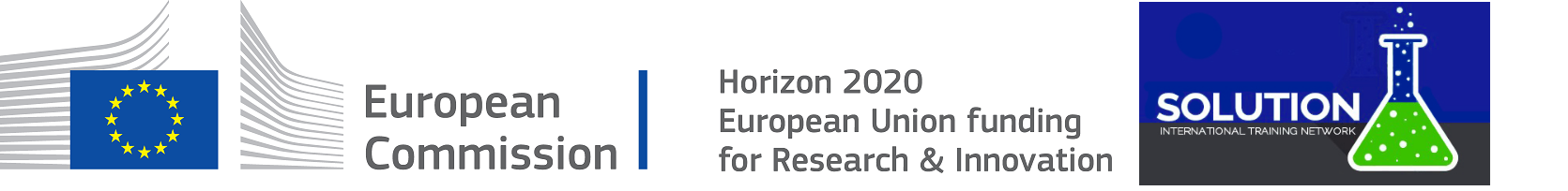}
\end{center}
\printbibliography
%
%
\end{multicols}
\end{document}